\pdfoutput=1

\documentclass{aastex62}

\usepackage{amsmath}
\usepackage{dsfont}
\usepackage{verbatim}
\usepackage{bm}
\usepackage[linesnumbered]{algorithm2e} 

\newcommand{\norm}[1]{\left\lVert#1\right\rVert}

\DeclareMathOperator*{\argmax}{arg\,max}
\DeclareMathOperator{\sech}{sech}

\graphicspath{{./}{figures/}}

\shortauthors{Dupuis et al.}

\begin{document}


\title{Characterizing magnetic reconnection regions using Gaussian mixture models on particle velocity distributions}

\correspondingauthor{Romain Dupuis}
\email{romain.dupuis@kuleuven.be, dupuis.ro@gmail.com}

\author[0000-0002-7976-1034]{Romain Dupuis}
\affil{ Center for mathematical Plasma Astrophysics,
  KU Leuven,
  Celestijnenlaan 200B, bus 2400, 
  B-3001 Leuven, Belgium
}

\author{Martin V. Goldman}
\affiliation{University of Colorado,
 Boulder, CO 80309, USA \\}

\author{David L. Newman}
\affiliation{University of Colorado, 
 Boulder, CO 80309, USA \\}

\author[0000-0003-1320-8428]{Jorge Amaya}
\affil{ Center for mathematical Plasma Astrophysics,
  KU Leuven, 
  Celestijnenlaan 200B, bus 2400, 
  B-3001 Leuven, Belgium 
}

\author[0000-0002-3123-4024]{Giovanni Lapenta}
\affil{ Center for mathematical Plasma Astrophysics, 
  KU Leuven,
  Celestijnenlaan 200B, bus 2400, 
  B-3001 Leuven, Belgium  
}



\begin{abstract}

We present a new method based on unsupervised machine learning to identify regions of interest using particle velocity distributions as a signature pattern. An automatic density estimation technique is applied to particle distributions provided by PIC simulations to study magnetic reconnection. The key components of the method involve: i) a Gaussian mixture model determining the presence of a given number of subpopulations within an overall population, and ii) a model selection technique with Bayesian Information Criterion to estimate the appropriate number of subpopulations. Thus, this method identifies automatically the presence of complex distributions, such as beams or other non-Maxwellian features, and can be used as a detection algorithm able to identify reconnection regions. The approach is demonstrated for a specific double Harris sheet simulations but it can in principle be applied to any other type of simulation and observational data on the particle distribution function.

\end{abstract}



\section{Introduction}
Plasmas are host to a complex mixture of interacting processes. Determining what process happen in a given location is often a challenge. When plasmas are modelled with a kinetic description, the researcher is confronted with a 6D data set. In Particle-In-Cell (PIC) methods, this information comes as a collection of hundreds or thousands of particles per cell with a total data size nowadays reaching a totality close to trillions of particles. Often the analysis focuses only on the electromagnetic fields and the moments of the particle distribution that are 3D manageable data sets. However, this leaves out the richest part of the simulation information: the particles. We consider here an approach based on the Gaussian mixture model to extract automatically information from the particle distribution without requiring human intervention.  

To demonstrate the approach we consider the often-studied case of a plasma undergoing magnetic reconnection. Magnetic reconnection \citep{gonzalez2016magnetic} plays a crucial role in collisionless plasmas. By breaking down the frozen-in magnetic fields, the magnetic field energy is converted into kinetic energy, thermal energy, and particle acceleration energy. This process appears fundamental in the transport mechanism and it represents one of the most important sources of particle acceleration in space. Magnetic reconnection can occur at various scales and locations such as: in laboratory plasma~\citep{yamada2014conversion}, in plasma turbulence~\citep{haynes2014reconnection}, or in the magnetotail~\citep{eastwood2013energy}. Therefore, the magnetic reconnection has been studied in many ways, including in-situ measurements with the Magnetospheric Multiscale (MMS) mission~\citep{burch2016magnetospheric} and numerical simulations~\citep{goldman2016can, hesse2014electron}. 

In this paper, we are interested in automatically characterizing reconnection regions from particle distributions.  We investigate this in a well documented case, that of two-dimensional collisionless PIC simulations. The literature of the last 20 years has numerous examples of variations of 2D reconnection setups, from simple Harris initializations \citep{birn2001geospace} to more realistic equilibria \citep{sitnov2013spontaneous} based on analytical models \citep{lembege1982stability} or on global MHD simulations \citep{ashour2015multiscale}. In this context, we can more readily interpret the result of the new diagnostic presented here using the knowledge-base of 2D PIC simulaitons accured in the recent past. Future work will then expand the use of the new diangostics to less documented cases such as turbulent reconnection in 2D and 3D and to distributions obtained from in situ space missions.  


The particle-scale kinetic physics, and in particular the electron-scale, has  recently received renewed interest  in the context of the magnetic reconnection thanks to MMS mission. In particular, electron distributions have shown to be good indicators for magnetic reconnection. For instance, exhaust electrons can give rise to highly structured anisotropy when the reconnection rate achieves its maximum~\citep{shuster2014highly}. Crescent-shaped distributions can be detected near the electron stagnation point  for asymmetric reconnection~\citep{burch2016electron} as indication of the presence of meandering orbits \citep{bessho2016electron}. Meandering orbits and crescents are observed also in other regions around a reconnection regions, such as in the proximity of the separatrix of asymmetric reconnection but also in symmetric reconnection \citep{lapenta2017origin,egedal2016spacecraft}. Triangular shapes have also been observed in the vicinity of the X line  within the electron diffusion region for weak guide fields~\citep{shuster2015spatiotemporal}. In presence of  magnetic islands, specific distributions can be present for each regions, such as flat-top or crescent-shaped distributions~\citep{cazzola2016electron}.
Thus, while distributions provide a richer insight on the local physics than the local fields and moments,  it seems clear that a unique specific distribution cannot be used as a signature for reconnection as it does not reflect the phenomenon for all the possible external conditions. For this reason, developing a detection algorithm based on machine learning techniques and able to detect non-Maxwellian features is especially desirable. Such methods can detect complex shapes from the analyze of the electron velocity distributions. Moreover, they could be coupled with other more classical detection methods based on field quantities, such as agyrotropy~\citep{aunai2013electron}.

Machine learning is increasingly being used in various fields related to physics. For instance, particle physics research has been very active and precursor in applying modern data analysis techniques on their problems. Machine learning, and even deep learning, are already established tools to analyze data. A whole work-flow has been developed to process the petabytes of data generated by the Large Hadron Collider: data reduction supports real-time analysis and data storage, boosted decision trees help for Higgs boson discovering, convolutional neural networks allow to reduce the noise from experiments, and recurrent neural networks identify quarks~\citep{radovic2018machine}. 
In space science and plasma physics, supervised machine learning has already been widely used \citep{camporeale2018machine}. Algorithms based on this approach aim at finding a relationship between input features and outputs. They can forecast geomagnetic indices~\citep{gruet2018multiple}, flares~\citep{florios2018forecasting}, and coronal mass ejections~\citep{bobra2016predicting}. They can classify solar wind with Gaussian processes in four categories defined by the solar origin of the wind~\citep{camporeale2017classification}. Moreover, regression techniques can accelerate the simulation of turbulence tokamak transport code to allow real-time analysis~\citep{citrin2015real}. A three-dimensional convolutional neural network (3D-CNN), trained on  human labelled  examples is able to predict plasma regions with 99\% accuracy \citep{olshevsky2019automated}. However, supervised learning algorithms would need a  database of magnetic reconnection-related distribution functions to be efficient. Building such a database is inconvenient and human labour-intensive, therefore other approaches can be considered.

Unsupervised learning, another type of machine learning techniques, extracts hidden structures and patterns from data without any corresponding pre-labelled target values. There is no more a mapping between inputs and outputs, as it was the case for supervised learning. Unsupervised learning encompasses mainly dimensionality reduction, clustering, generative modeling, and density estimation. Such techniques have been used very little in space weather and plasma physics \citep{Bishop2006}. Only few examples of unsupervised learning can be found in the literature. For instance solar wind plasma has been divided into different categories using a k-means clustering algorithm~\citep{heidrich2018solar}. The authors proposed a physical interpretation to these categories and suggest specific features for future solar wind categorization. In a context closer to this paper, a wavelet-based density estimation has been used to post-process discrete particle data from PIC simulations to estimate distribution functions in order to separate the relevant information from the noise~\citep{del2010wavelet}. 

We are particularly interested in density estimation techniques, approximating the probability density function from the data~\citep{Bishop2006}. They are especially promising in identifying specific physical regimes, such as magnetic reconnection. In particle physics, authors have proposed to use density estimation to detect the presence of new physics events in the data~\citep{albertsson2018machine}. It gave us the idea of applying such techniques on reconnection simulations with the goal of detecting specific particle distributions, such as beams or non-Maxwellian distributions, which could be used as a magnetic reconnection signature. Therefore, the method proposed in this paper contributes to improve the potential connections between machine learning and plasma physics in general by providing a relevant illustration of magnetic reconnection detection.

This article is organized as follows: Section 2 introduces several classical reconnection signatures, Section 3 gives details on various velocity distributions related to reconnection, Section 4 presents the strategy to identify reconnection regions, Section 5 describes the PIC simulations, and the results are discussed in Section 6. 

\section{Identifying reconnection signatures}
As a specific example for untrained automatic detection of features in a plasma we select the process of magnetic reconnection. Magnetic reconnection is associated to the presence of an electron diffusion region (EDR), which modifies the magnetic field due to the generation of dissipative electric fields~\citep{birn2007reconnection}. As this region is very small and localized, its precise detection is very hard, especially for spacecraft measurements. For this reason, indirect signatures of reconnection are also of main interests, such as the presence of fast flows or plasma heating~\citep{hesse2014electron}. A wide range of signatures has been highlighted in the literature, using different source of data. We refer the readers to a recent review of all the signatures of reconnection proposed in recent years~\citep{goldman2016can}. 

These measures can be organized in two groups. The first is based on field quantities: for example, the detection of magnetic nulls \citep{fu2015find} and magnetic skeletons~\citep{haynes2007magnetohydrodynamic} or the explicit violation of magnetic flux conservation~\citep{newcomb1958motion,vasyliunas1975theoretical,hesse1988theoretical}. The second category uses the moments of the plasma species: for example the relative drift between the plasma and the field lines (slippage) or the energy dissipation measured on the electron frame~\citep{zenitani2011new}. As a specific example of this category, we will use here the second order moment, the pressure tensor, that defines the agyrotropy, proposed to identify reconnection sites~\citep{scudder2008illuminating}. Indeed, nongyrotropic velocity distributions are expected to play a key role during the magnetic reconnection process as significant electron pressure nonygrotropies can provide the reconnection electric field~\citep{hesse2014electron}. Usually, the different methods quantify the deviations from symmetry for the pressure tensor. However, this concept has lead to different definitions~\citep{scudder2008illuminating, aunai2013electron, swisdak2016quantifying} with varying degrees of success to identify reconnection. In the present paper we rely specifically on the measure of agyrotropy called $Q$~\citep{swisdak2016quantifying}:
\begin{equation}
    Q = \frac{P_{12}^2 + P_{13}^2 + P_{23}^2}{P_\parallel + 2 P_\bot },
\label{eq:mesure_gyro}
\end{equation}
where $P_\parallel$ and $P_\bot$ are the diagonal terms of the tensor and $P_{12}$, $P_{13}$, and $P_{23}$ are the sub and upper diagonal terms of the symmetric tensor. The agyrotropy measure is equal to 0 for gyrotropic tensors while $Q$ is equal to 1 for maximal deviations. 

In the present work we present a third new approach at detecting reconnection: using directly the distribution function and measuring its complexity via the Gaussian mixture model, an unsupervised machine learning approach largely used in other fields~\citep{Bishop2006}.

\section{Fitting particle velocity distributions}
Fitting distributions to a sample of data is the process of choosing a probability distribution modelling a data set and estimating the associated parameters. The selection of a correct distribution function must take into account various parameters involving mathematical and physical arguments. Is the distribution unimodal or multimodal? Does the phenomena show symmetric or skewed behavior? Can we derive specific bounds for the distributions? Answering to these questions will guide the choice of the model. 

Various distributions have been used to fit plasma particle velocity distributions. The most common model in space plasmas is the Maxwellian and the bi-Maxwellian, taking into account temperature anisotropy. The classical Maxwellian distribution shows good results by describing velocity distribution in low-energy regions, in particular for ions~\citep{gruntman1992anisotropy, kasper2006physics}. However, the plasma velocity distributions exhibit non-Maxwellianities for suprathermal regions, where the distribution is governed rather by power law tails. Thus, the Kappa distribution (also called generalized Lorentzian) has been proposed to describe both low-energy Maxwellian cores and suprathermal tails~\citep{vasyliunas1968survey, summers1991modified}. Kappa distributions have gained an important notoriety in numerous studies in space plasmas~\citep{hellberg2002generalized, pierrard2010kappa, livadiotis2013understanding, ogasawara2013characterizing, lazar2018temperature, livadiotis2018generation}. One can note that when the spectral index kappa increases towards infinity, the Kappa distribution tends to a Maxwellian.

Previous studies have shown examples of electron velocity distributions fitting using one dimensional cut~\citep{pulupa2014spin} or two dimensional distributions~\citep{wilson2019electron} for the solar wind. In the latter paper, the best approximation is built as a sum of three densities for the cold dense core, the hot halo and the beam. Each component is fit by chosen among a list of potential distributions and optimizing the associated parameters. This kind of approach allows to provide a physical interpretation to the main distributions of each components. However it relies on a strong physical knowledge such as the list of potential distributions or the range of variations for all distribution parameters. Such detailed knowledge is not necessary available or reliable for all physical phenomena or locations.~\citet{souza2018classification} use an automatic clustering method, called self-organizing map, to organize pitch angle-resolved particle flux data collected in the outer Van Allen belt region into different categories. As regards reconnection, we described previously that various distribution shapes could be observed near reconnection sites, such as crescent-shapes or triangles, and their presence may depend on various conditions. Their discovery may sound recent and maybe other distributions exist and have not yet been discover. An ideal algorithm must therefore not rely on specific set of distributions for the detection of reconnection. 

Density estimation techniques aim at building a model of a non-observable probability density function by observing a set of data points. They appear therefore as a potential candidate to automatically fit complex distribution functions. We expect to identify particles distributions with specific shapes, such as beams or non-Maxwellian features in order to relate them with reconnection sites. A  growing interest for such methods has been observed in astronomy~\citep{ivezic2014statistics_chap6}. There are two main kind of density estimations: parametric and nonparametric methods. The first one represents the natural approach where the distribution is estimated by fitting the parameters of a given model to the data. For instance, a Gaussian distribution can be locally approximated by a second order polynomials. \citet{ni2015variability} fit electron pitch angle distributions using $sin^N(\alpha)$ functions where $\alpha$ is the local particle pitch angle and $N$ the power law. A very popular method, called Gaussian Mixture Model, fits the data with a sum of Gaussian distributions~\citep{Bishop2006}. On the other hand, nonparametric methods try to make as few assumptions as possible, mainly by working with infinite-dimensional models. One of the simpler nonparametric density estimators is the histogram which splits the support of the distribution into bins and then the value of the function is defined as the number of samples falling into that bin. A very popular method in machine learning, called Kernel Density Estimation (KDE), proposes a more general approach by convolving the data with a smooth kernel function~\citep{sheather2004density}. However, for the two methods, a specific issue arises as the width of the kernel (or the size of the bin) must be chosen. If this value is to small, a noisy function is observed as randomness in the signal is highlighted. If the value is too large, modes are smoothed out and important structures are obscured. Several strategies have been proposed to determined this parameter, such as cross-validation or plug-in methods~\citep{heidenreich2013bandwidth}. Finally, one may note that nonparametric density estimators are inherently linked to the data. Their expressions are defined with the data while once the parameters of a parametric methods have been tuned, the data can be throw away. Therefore the definition of the parametric estimator is conserved if the data are lost. This is not the case of nonparametric models. 

\section{Detection algorithm}
In this section, a detection algorithm based on the Gaussian Mixture Model (GMM) is presented. Parametric methods were preferred over nonparametric ones as they provide an easier interpretability. After introducing the main mathematical derivations of the GMM, the selection of the number of components is detailed and two specific metrics based on thermal energy are defined.

\subsection{Density estimation with Gaussian mixture models}
 The mixture model is defined as a weighted sum of given densities with unknown parameters. The most common density is the Gaussian density as it ensures a closed formalism for the determination of the parameters and limits the computation to the means and the covariances. Moreover, the second reason is that Gaussian density can be considered as a reasonable assumption for the density when no prior information is available for the probability density function. A general technique for finding the unknowns parameters consists in maximizing the likelihood function with the expectation-maximization (EM) algorithm. This approach is detailed below. 

As regards the mathematical formalism, the random variable $\bm{x}$ associated to the observations is assumed to be written as a linear superposition of $K$ multivariate Gaussians:
\begin{equation}
p(\bm{x}| \bm{\Phi}) = \sum_{k=1}^K w_k \, \mathcal{N}(\bm{x}| \bm{\theta_k}).
\label{eq:gmm}
\end{equation}
The normal distribution $\mathcal{N}$ is parameterized by the mean $\bm{\mu_k}$ and covariance matrix $\bm{\Sigma_k}$ of the $k$-th mixture regrouped in $\bm{\theta_k}$ and the proportion $w_k$. All the parameters of the Gaussian mixture model are regrouped in the mixture parameter $\bm{\Phi} = [w_1, \ \cdots, \  w_q, \ \bm{\theta_1}, \ \cdots, \  \bm{\theta_K}]$. The Python package Scikit-learn is used to perform all the computations~\citep{pedregosa2011scikit}.

\subsubsection{Unobserved latent variables}
A K-dimensional binary random latent variable $\bm{z}\in \mathds{R}^K$ is introduced such as a particular component $z_k$  of $\bm{z}$ is equal to 1 and all other elements are equal to 0, meaning that $z_k$ satisfies  $z_k \in \{0,1\}$ and $\sum\nolimits_{k=1}^K z_k =1$.  In particular, the $k$-th component is $1$ if the observation of $\bm{x}$ is generated from the $k$-th Gaussian such as the marginal distribution over $\bm{z}$ is directly related to the mixture proportion as $p(z_k =1) = w_k$.

We want to rewrite the definition of the Gaussian mixture in~\autoref{eq:gmm} by introducing the unobserved latent variable $\bm{z}$. The distribution of this latter can be expressed as:
\begin{equation}
    p(\bm{z}) = \prod_{k=1}^{K}w_k^{z_k}.
\end{equation}
Moreover, the conditional distribution of $\bm{x}$ given the latent variable $\bm{z}$ is straightforward:
\begin{equation}
    p(\bm{x} | \bm{z}) = \prod_{k=1}^{K}\mathcal{N} (\bm{x} | \bm{\theta_k})^{z_k}.
\end{equation}
The joint distribution of $\bm{x}$ and $\bm{z}$ is expressed with the product rule:
\begin{equation}
    p(\bm{x}, \bm{z}) = p(\bm{z})p(\bm{x} | \bm{z}) =  \prod_{k=1}^{K}w_k^{z_k} \prod_{k=1}^{K}\mathcal{N} (\bm{x} | \bm{\theta_k})^{z_k}.
\end{equation}
Finally, the marginal distribution is integrated over $\bm{z}$ and we find the same expression as in~\autoref{eq:gmm}:
\begin{equation}
    p(\bm{x}) = \sum_{\bm{z}} p(\bm{z}) p(\bm{x}, \bm{z}) = \sum_{k=1}^K w_k \mathcal{N} (\bm{x} | \bm{\theta_k})^{z_k}.
\end{equation}
This new expression of the Gaussian mixture involves now the latent variable $\bm{z}$ and will be very useful to compute all the parameters of the mixture as we can now work with the joint distribution function instead of the marginal distribution. Moreover, each observation $\bm{x_i}$ is now associated to a specific value of the latent variable $z_i$.

Let now introduce the condition probability of $\bm{z}$ given the observation $\bm{x}$ and called $\gamma$:
\begin{equation}
\begin{aligned}
    \gamma(z_k) := p(z_k = 1 | \bm{x})  &= \frac{p(z_k = 1) p(\bm{x} | z_k=1)}{p(\bm{x})}\\
                                        &= \frac{w_k \mathcal{N}(\bm{x} | \bm{\theta_k} )}
                                        {\sum_{j=1}^K w_j \mathcal{N}(\bm{x} | \bm{\theta_j} )}.
\end{aligned}
\label{eq:gmm_gamma}
\end{equation}
It can also be viewed as the responsibility that component $k$ takes for ‘explaining’ the observation $\bm{x}$~\citep{Bishop2006}.

\subsubsection{Maximum likelihood with Expectation Maximization algorithm}
Let assume we have a set of observations $\bm{x_1}, \cdots, \bm{x_n}$ from the variable $\bm{x}$. They form the matrix of the training set $\bm{X} \in \mathds{R}^{n\times p}$. The same procedure is applied to build the matrix $\bm{Z}\in \mathds{R}^{n\times p}$ associated to the latent variable values.
The log likelihood $l$ of the Gaussian mixture can be expressed from the observations:
\begin{equation}
    l(\bm{\phi} | \bm{X}, \bm{Z}) = \sum_{i=1}^n \ln \left[ \sum_{k=1}^K w_k \mathcal{N}(\bm{x_i} | \bm{\theta_k} )  \right].
    \label{eq:gmm_likelihood}
\end{equation}
Maximizing this expression appears as a complex problem due to the summation inside the logarithm. Two main approaches exist to solve this maximization problem: classical gradient-descent or expectation-maximization (EM) algorithm~\citep{Bishop2006}. We will focus on the latter.

The mixture parameters regrouped in $\bm{\Phi}$ are estimated iteratively using an Expectation Maximization algorithm (EM algorithm)~\citep{Dempster1977}. Let set the gradient of the likelihood expression in~\autoref{eq:gmm_likelihood} to zero with regards to: i) the mean $\mu_k$, ii) the covariance matrix $\Sigma_k$, and iii) the mixture proportion $w_k$ (coupled with a Lagrange multiplier to take into account the constraint $\sum\nolimits_{k=1}^K z_k =1$). After several derivations detailed in~\citet{Bishop2006} and by writing $\gamma(z_{ik})$ the specific value of the responsibility for a given observation $\bm{x_i}$, we end up with the three expressions:
\begin{equation}
\bm{\mu_k} = \frac{\sum\limits_{i=1}^n
\gamma(z_{ik})\bm{x_i}}{\sum\limits_{i=1}^n \gamma_(z_{ik})}, \ \forall k
\in [1,  \cdots, K],
\label{eq:gmm_mean}
\end{equation}

\begin{equation}
\bm{\Sigma_k} = \frac{\sum\limits_{i=1}^N
\gamma(z_{ik})(\bm{x_i}-\bm{\mu_k})(\bm{x_i}-\bm{\mu_k})^{T}
}{\sum\limits_{i=1}^N \gamma(z_{ik})}, \ \forall k
\in [1, \cdots, K],
\label{eq:gmm_cov}
\end{equation}

\begin{equation}
w_k = \frac{1}{n}\sum_{i=1}^n \gamma(z_{ik}), \ \forall k
\in [1,  \cdots,K].
\label{eq:gmm_mixture}
\end{equation}

However, these three expressions do not provide a closed-form solution for the parameters of the mixture due to the complex relationship between them and $\gamma$ expressed in~\autoref{eq:gmm_gamma}~\citep{Bishop2006}. For this reason, a simple iterative scheme for finding a solution to
the maximum likelihood problem (EM algorithm) is used for this particular case of the Gaussian mixture model. The EM algorithm is split into two steps: the expectation step (also called E step) and the maximization step (also called M step).

In the E step, the posterior probability of $\bm{z}$ (the responsibility) is computed from the current value of the parameters. Then, in the M step, all the parameters are re-estimated by using the previously computed posterior probability. The algorithm can be written with the following steps~\citep{Bishop2006}:
\begin{enumerate}
    \item [0] Initialize the parameters of the mixture $\bm{\Phi}$: means $\bm{\mu_k}$, covariances $\bm{\Sigma_k}$, and mixing coefficients $w_k$.
    \item [1] \textbf{E-step}: Compute the responsibility $\gamma(z_{ik})$ from~\autoref{eq:gmm_gamma}.

    \item [2] \textbf{M-step}: Re-estimate the parameters $\bm{\mu_k}$, $\bm{\Sigma_k}$, and $w_k$ from respectively~\autoref{eq:gmm_mean}, \ref{eq:gmm_cov}, and \ref{eq:gmm_mixture} using the current responsibilities $\gamma(z_{ik})$. 
    
    \item[3] Check the convergence. If not, return to 1.
\end{enumerate}


\subsection{Model selection}
The number of Gaussians $K$ usually acts as an input to the GMM algorithm. This value may be specified by the user at the beginning of the algorithm or it can be estimated by analyzing the data. Several methods are proposed in the literature: cross validation, elbow method, information criterion, etc~\citep{Bishop2006}. The general idea is to define an estimator related to the relative quality of the Gaussian mixture for a given set of data. Information criteria represent good candidates as they give a trade-off between the goodness of fit and the complexity of the model. The two main estimators are called Akaike Information Criterion (AIC) and Bayesian Information Criterion (BIC)~\citep{anderson2004model}: 
\begin{equation}
\begin{aligned}
AIC &= 2k - 2\ln(L) \\
BIC &= \ln(n)k - 2\ln(L) \\
\end{aligned}
\label{eq:aic_bic}
\end{equation}
where $k$ is the number of parameters to estimate in the model and $L$ the likelihood. BIC penalizes more the model complexity than AIC. However, AIC and BIC performances depend on the nature of the data generating model: sample size, complexity or the model, whether the true model is contained in the model set or not, etc~\citep{anderson2004model}. As data from simulations may be noisy and the number of particles is significant, BIC has been preferred in this work to automatically select the number of components of the mixture.

Nevertheless, the physical meaning of the number of components $K$ and the parameters associated to each Gaussian must be analyzed carefully as they must not be necessary interpreted as specific beams or electron populations. Indeed, if the data show complex shapes or are not near Gaussian, the number of components $K$ does not correspond to the number of different populations~\citep{ivezic2014statistics_chap6}. For instance, a flat-top distribution is approximated by several Gaussians but each component is needed to approach the broad mode of the distribution. A Kappa distribtion can also be represented by a central Gaussian centered around the mode plus another Gaussian with a very large width to fit the wide tail, thus 2 Gaussians are needed for a single population. Moreover, as presented previously, BIC is sensitive to various parameters: the data themselves and the sample size. For instance, if the source of the data does not change but the number of samples increased, the resulting number of components may also change. However, BIC is still a efficient criterion to provide a statistical analysis based on underlying properties of the data. It can help to detect important variations in the distribution. Another strategy consists in fixing the number of components to a high value in order to improve the fit for very complex distributions which can show poor results for a small number of components. In this case, GMM is very close to a nonparametric density estimation method, such as KDE. Such strategy is illustrated in the~\autoref{app:fixed_number}.

\subsection{Thermal energy variation}
As the particle distributions are approximated by sums of Gaussians instead of a single Maxwellian, it is interesting to analyze the variation of the thermal velocity for these two representations. The thermal energy for a single velocity distribution is given by its variance. The straight measure of thermal energy based on the moment of the whole distribution is:
\begin{equation}
E_{thermal} = \frac{1}{N_p}  \sum_{i=1}^3  \left[ \sum_p \left( \bm{V_p} - \langle \bm{V_p} \rangle \right)^2\right]_i, \text{ with } \langle \bm{V_p} \rangle = \sum_p \frac{\bm{V_p}} {N_p}.
\label{eq:th_energy}
\end{equation}
The variance $(\sigma^2)^{(K)}$ for $K$ multiple Maxwellians is given by:
\begin{equation}
    (\sigma^2)^{(K)} = \sum_{i=1}^3  \left[ \sum_{k=1}^K w_k^2 \left( \bm{\sigma_k} \right)^2 + \sum_{k=1}^K w_k \left( \bm{\mu_k} \right)^2 - \left(\sum_{k=1}^K w_k (\bm{\mu_k}) \right)^2  \right]_i.
    \label{eq:full_var}
\end{equation}
The first term can be interpreted as the mixture of the variances and is related to the thermal energy per unit mass of the mixture. Therefore, it is written as the thermal energy (per unit mass) of the $K$ multiple Maxwellians:
\begin{equation}
E_{thermal}^{(K)} = \frac{1}{2} \sum_{i=1}^3 \sum_{k=1}^K w_k^2 \left[ \bm{\sigma_k}^2 \right]_i.
\end{equation}
The thermal energy ratio $E_{drop}$ is derived to compute the reduction in thermal speed for the particles, aiming to distinguish heating from accelerating particles into beams. It measures the ratio between the mixture of the variance and the variance of the velocity distribution:
\begin{equation}
    E_{drop} = \frac{E_{thermal}^{(K)}}{E_{thermal}}.
\label{eq:energy_drop}
\end{equation}
This metric is defined to always be below $1$. Low values indicate that the thermal energy of the mixture is much smaller than the thermal velocity computed directly from the definition, suggesting that the second order moment of the overall distribution is not a good indicator of the conditions present. An extreme example is that of two cold beams which individually have zero thermal spread and only a relative mean velocity but when taken together appear as a broad thermal spread. This measure identifies these conditions, spotting distributions characterized by interpenetrating beams.

The last two terms of~\autoref{eq:full_var} can be read as the deviation of each mean compared to the overall mixture mean:
\begin{equation}
E_{dev}^{(K)} = \sum_{i=1}^3  \left[ \sum_{k=1}^K w_k \left( \bm{\mu_k} \right)^2 - \left(\sum_{k=1}^K w_k (\bm{\mu_k}) \right)^2  \right]_i.
\end{equation}
This deviation is always positive as it corresponds to a weighted variance. This is the thermal energy of the center of all beams, measuring the distance between them. A second metric $E_{dev}$, called thermal velocity deviation, defines the ratio between the velocity deviation for the mixture and the classical thermal velocity of the distribution:
\begin{equation}
    E_{dev} = \frac{E_{dev}^{(K)}}{E_{thermal}}.
\label{eq:energy_deviation}
\end{equation}
This strictly positive quantity allows to interpret the different mixtures. High values mean the components are widely separated and presumably have a distinct identity and perhaps origin \citep{eastwood2015ion}. Small values point to mixtures of components close to each other and perhaps carry less meaningful separation.

\subsection{General procedure}
\label{sub:general_procedure}
\begin{figure}
    \centering
    \includegraphics[width=0.5\linewidth]{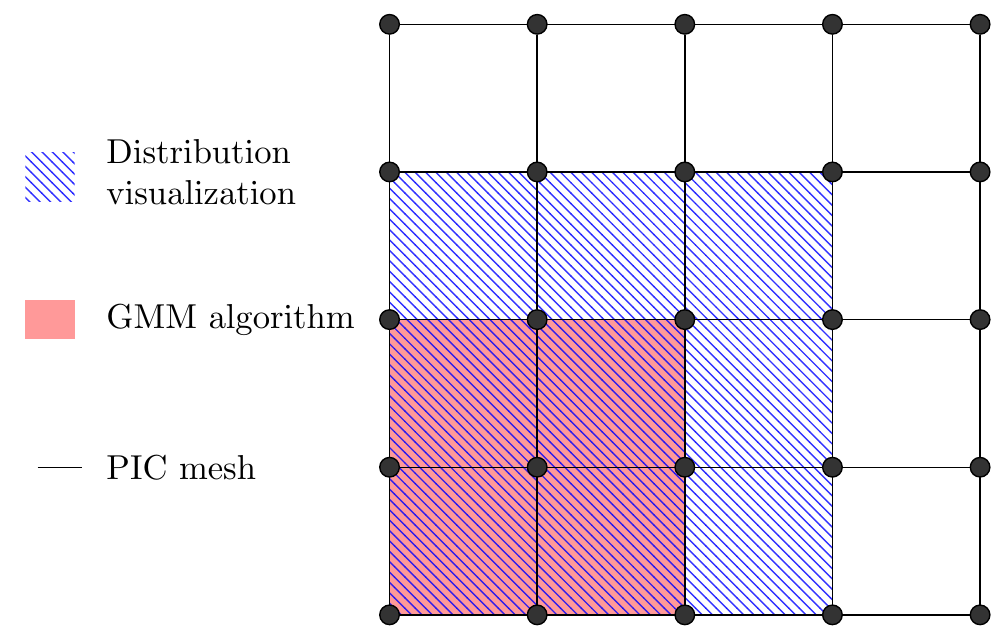}
    \caption{Illustration of the different levels of granularity between the PIC cells, the GMM algorithm cells, and the distribution visualization regions. They may be at different levels due to few number of particles.}
    \label{fig:mesh_level}
\end{figure}

The detection algorithm has been performed on a square window with a size of $r$ by $r$ cells. This set of cells is merged in each direction to build the computational domain. Using large windows aims at providing enough particles to the algorithm. It ensures that the model selection by BIC is less sensitive to the number of particles and does not necessary favor very simple models with few numbers of components with regards the real complexity of the data. The~\autoref{fig:mesh_level} illustrates the different levels between this square window, the PIC mesh, and the visualization regions for the distributions. We are limiting the maximum number of components to $6$ for the BIC optimization. The general procedure is given by the following algorithm:
\begin{algorithm}[H]
 \KwData{particles coordinates $\bm{X}$ and velocities $\bm{V}$, mesh $\Omega$ with $n_x$ cells by $n_y$ cells}
 \KwResult{distribution function $f_i$, $E_{drop}$, and $E_{dev}$ for each subset $i$}
 \Begin{
 merge cells to increase the number of particles by subset, new mesh $\Omega'$ is of size $\frac{n_x}{r} \times \frac{n_y}{r}$\;
 \For{$i \in \Omega'$}{
  \For{$k = 1$ to $k=6$ }{ 
  $GMM_k = GMM(\bm{V_i}; k)$\;
  $BIC_k = BIC(GMM_k)$\;
  }
  $K = \argmax_{k \in [1,6]} BIC_k$\;
  $f_i = GMM_K$\;
  $\text{compute }E_{drop} \text{ and } E_{dev}$\;
 }
 }
 \caption{Detection algorithm}
\end{algorithm}

\section{PIC simulations}

The simulations are performed with the fully kinetic massively parallel implicit moment method Particle-in-Cell code iPic3D~\citep{markidis2010multi, innocenti2017progress}. We are particularly interested in a double Harris sheet case where two well separated reconnection sites can be identified. Two different guide fields have been tested: a weak and a strong one. The simulation is 2.5D, thus all the vectors are considered as three dimensional, but their spatial variation is limited to a two dimensional plan independent of the dawn-dusk (Z) direction. All quantities are presented in normalized form.

PIC simulations of 2D Harris sheet~\citep{Harris1962} reconnection are paradigmatic in reconnection and are   therefore our choice in determining how the diagnostic presented here performs in this classic well known problem, often also referred to as GEM Challenge~\citep{birn2001geospace}. We consider here specifically the double Harris sheet case defined by:
\begin{equation}
B_x(y) = B_0  (-1 + \tanh(y-y_1) - \tanh(y-y_2)),
\end{equation}
with the location of the two current layers at $y_1=L_y/4$ and $y_1=3L_y/4$~\citep{wu2011effect}. 
Pressure balance is kept by a uniform temperature but a non uniform density:
\begin{equation}
n_s(y) = n_0  (-1 + \sech(y-y_1)^2 + \sech(y-y_2)^2) +n_b,
\end{equation}
where a background density equal to $n_b=n_0/10$ is added.
The equilibrium is defined by the thickness $L/d_i=0.5$ and with the parameters $m_i/m_e=256$, $v_{the}/c=0.045$, $T_i/T_e=5$. With these choices, the asymptotic in plane  field $B_0$  is set by the ratio $\omega_{ci}/\omega_{pi}=0.0097$ and  the peak Harris density $n_0=1$ is imposed by the normalisation used that results in the ion plasma frequency and ion inertial length to be unitary. 
The coordinates are chosen with  the initial  Harris magnetic field  along $x$ with size  $L_x=30d_i$, the initial gradients  along $y$ with $L_y=40d_i$. The third dimension, where the initial current and guide field are directed,  is  invariant. Periodicity is assumed in all directions. The Cartesian mesh has a size of $769\times1025$ and about $98.300.000$ particles are injected in the computational domain, representing approximately $125$ particles by cell. 
The particle distributions are analyzed in a frame of reference driven by the local magnetic field, in addition to the Cartesian system, as suggested in~\citet{goldman2016can}. The B-field-aligned basis is defined by the following three vectors:
\begin{equation}
\begin{aligned}
\bm{e_{\parallel}} &:= \bm{\widehat{B}}, \text{ where } \bm{\widehat{B}} = \frac{\bm{B}}{\norm{\bm{B}}}\\
\bm{e_{\bot 1}} &:= \bm{\widehat{B}} \times \bm{e_z}\\
\bm{e_{\bot 2}} &:= \bm{\widehat{B}} \times \bm{e_{\bot 1} } = -\widehat{B}^2 \bm{e_z} + (\bm{e_z} .  \bm{\widehat{B}})  \bm{\widehat{B}} .
\end{aligned}
\end{equation}
Therefore, $\bm{e_{\parallel}}$ is parallel to the total magnetic field, $\bm{e_{\bot 1}}$ is in the reconnection x–y plane, perpendicular to in-plane magnetic field lines, and $\bm{e_{\bot 2}}$ is in the -z-direction for magnetic field with small z-component. 

\section{Results}
The automatic detection algorithm presented above can be used to spot magnetic reconnection and regions of interest. From the point of view of plasma physics, the decomposition in several Gaussians by the GMM algorithms can be considered as reasonable only if results are in agreement with other classical methods. For this reason, outcomes of the detection algorithm are compared with measures of agyrotropy considering the weak guide field case. The strong guide field is described in~\autoref{app:strong_guide}. As the Double Harris sheet case shows very similar behavior for the two layers, the results are only presented for the bottom layer. Moreover, as a significant number of particles is best suitable to train the density estimation model, the algorithm is performed on a coarser resolution. Each group of 4 cells by 4 cells ($r=4$) are merged into a square window.~\autoref{app:resolution_sensitivity} gives details to the sensitivity to this number of cells. Finally, only electron distributions are investigated in this paper.
\begin{figure} [ht]   
\gridline{
  \includegraphics[trim={0.55cm 3.75cm 1.25cm 3.85cm},clip,width=0.5\linewidth]{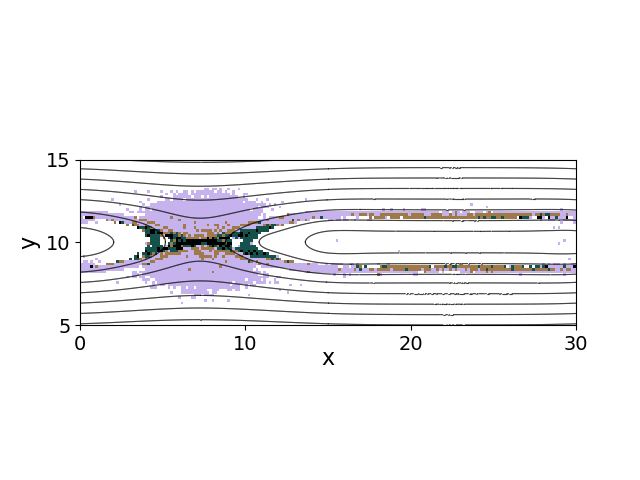}
    \includegraphics[trim={0.55cm 3.75cm 1.25cm 3.85cm},clip,width=0.5\linewidth]{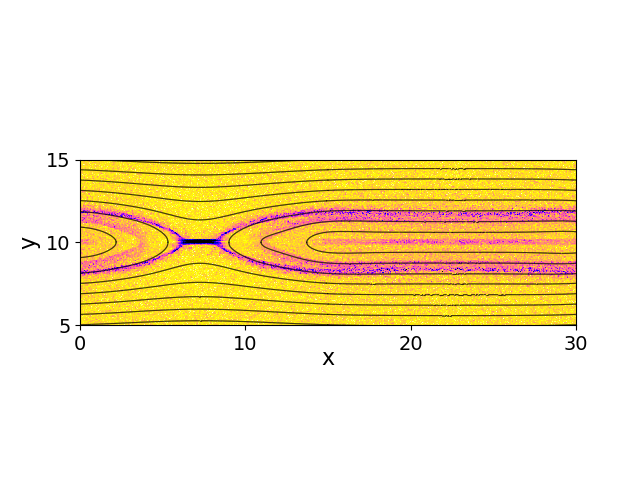}
}

\vspace{-0.35cm}

\gridline{
  \includegraphics[trim={0.55cm 3.75cm 1.25cm 3.85cm},clip,width=0.5\linewidth]{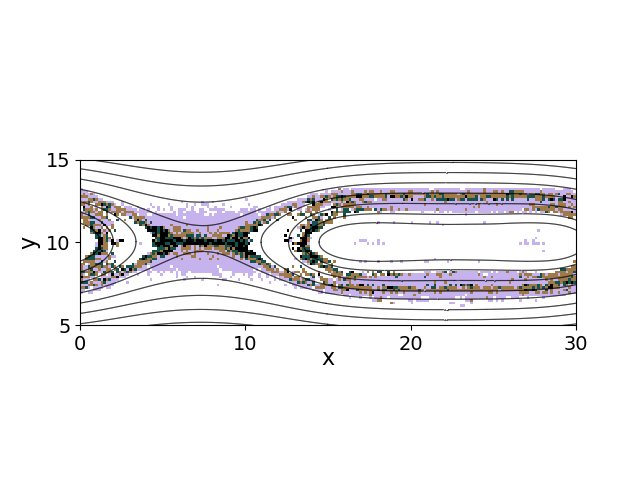}
   \includegraphics[trim={0.55cm 3.75cm 1.25cm 3.85cm},clip,width=0.5\linewidth]{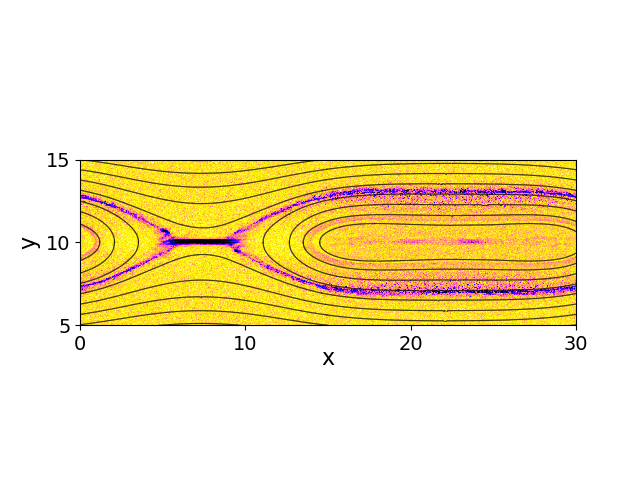}
}

\vspace{-0.35cm}

\gridline{
  \includegraphics[trim={0.55cm 3.75cm 1.25cm 3.85cm},clip,width=0.5\linewidth]{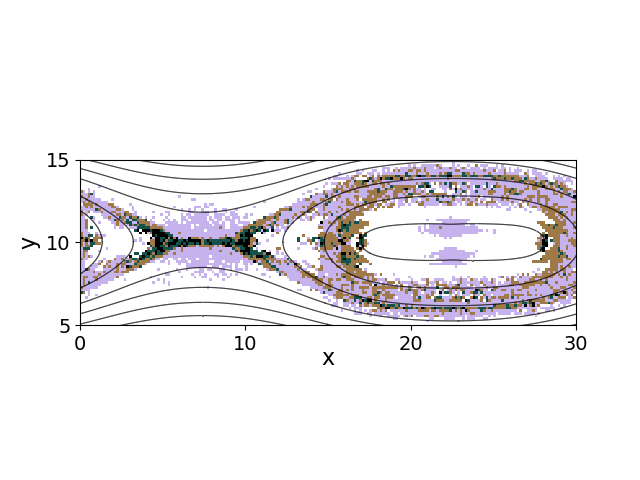}
    \includegraphics[trim={0.55cm 3.75cm 1.25cm 3.85cm},clip,width=0.5\linewidth]{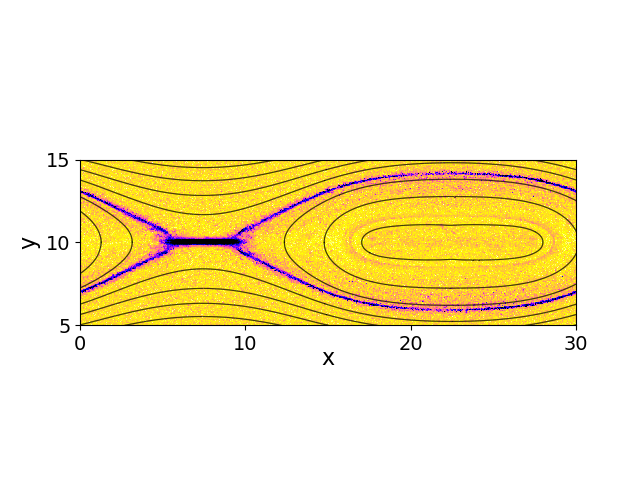}
}

\vspace{-0.35cm}

\gridline{
  \includegraphics[trim={0.55cm 1.5cm 1.25cm 3.65cm},clip,width=0.5\linewidth]{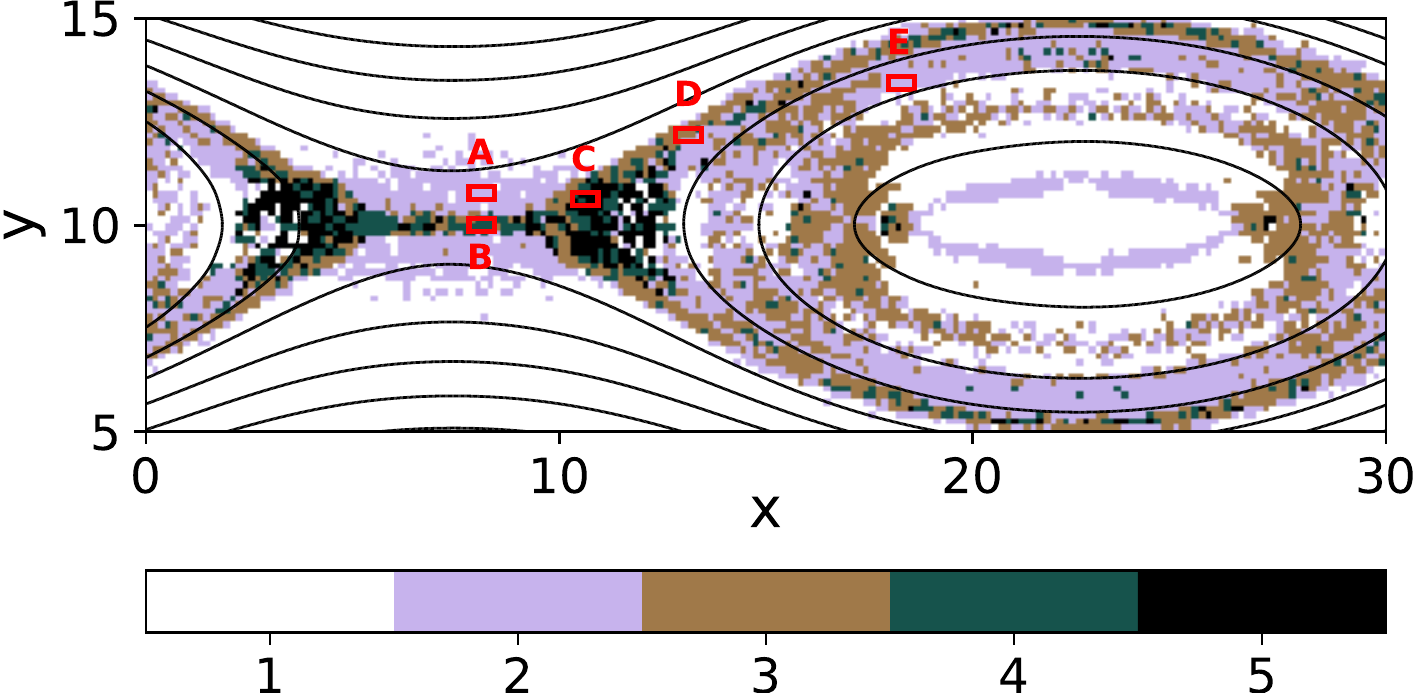}
    \includegraphics[trim={0.55cm 1.5cm 1.25cm 3.65cm},clip,width=0.5\linewidth]{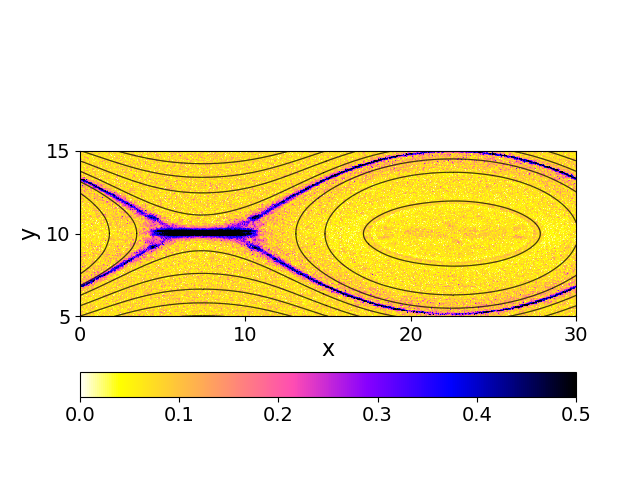}
}
\setlength{\belowcaptionskip}{-8pt}
 \caption{Magnetic reconnection detection for the Double Harris sheet case with a weak guide field at four different time steps, from top to bottom: $t=8,000$, $t=12,000$, $t=16,000$, and $t=20,000$. The left-hand column presents the number of components provided by the BIC optimization and the right-hand column shows the measure of agyrotropy $\sqrt{Q}$ defined in~\autoref{eq:mesure_gyro}. The red rectangles indicates the location where specific distributions are observed. They merge 4 GMM cells in the $x$ direction and 2 GMM cells in the $y$ direction.}
\label{fig:dh_result_weak}    
\end{figure}

\autoref{fig:dh_result_weak} compares the number of components identified by the detection algorithm on the left-hand column with the measure of agyrotropy on the right-hand column for various time steps. The objective is to highlight the behavior of the two quantities when the reconnection grows. Considering first the number of components, different structures are observed. Indeed, it can be clearly stated that not only the EDR is detected but a much wider panel of different regions, which are symmetric with respect to the central plane $y=10$. The algorithms seems to locate: inflows, ion and electron diffusion regions, outflow, and separatrix boundaries. Another striking result of the~\autoref{fig:dh_result_weak} is the capability of the algorithm to detect regions where the influence of the reconnection seems to be week, such as far upstream of the X line and near the O point. The noise of the PIC simulations is filtered out and unique distributions are successfully recognized. 
Starting with the first time step $t=8,000$, a large background tagged with $2$ components extends from $y\approx7$ to $y \approx 13$ and surrounds the EDR located at $x \approx 7$. This region may correspond to the ion diffusion region. The EDR is mainly composed by mixtures with $5$ and $4$ components, highlighting complex velocity distributions, while a transition region appears close to the EDR with $3$ components. Downstream from the EDR in the outflow, a C-shape structure can be noticed on each side, characterized by distributions with $4$ components connecting the EDR with the separatrix region. The latter is mainly composed of distributions with $2$ and $3$ components. 

As regards the three other time steps $t=12,000$, $t=16,000$, and $t=20,000$, they show very similar structures and behaviors. The size of the EDR tends to slightly increase over the time while the extend of the ion diffusion region reduces. The outflow region is still well identified and its location remains quite steady. The reconnection gives rise to a clear magnetic island on the right-hand side of the figure at these time steps. The thickness of the region around the O point tends to increase dramatically in the y direction when the reconnection grows. Several different distributions types can be observed, leading to a rather noisy mix with a background with $2$ components and some $3$ and $4$ components. Moreover, secondary structures gradually appears near the O point, creating a link between the bottom and the top layer of the island driven by distributions with $4$ components at $t=12,000$ and then $2$ and $3$ components later. It is important to note that no spatial constraints or correlations are imposed to the detection algorithm, thus all the structures identified by the BIC minimization may exist in the distributions. 

All the results provided by the detection algorithm are then compared to the values of agyrotropy depicted in the right-hand column in~\autoref{fig:dh_result_weak} for the same time steps. Few similarities are observed: the measure of agyrotropy clearly highlights the EDR for all time steps with peak values observed above $0.5$ and topological boundaries of the reconnection are also mapped, almost coinciding with the boundaries of the GMM algorithm with slight differences. However, different behaviors are exhibited compared to the detection algorithm. For instance, the region surrounding the EDR is not diagnosed by the agyrotropy as well as the outflow and inner structures around the O point. Small artifacts seem to be present within the topological boundaries but the background noise prevents them from being clearly identified. Indeed, the measure of agyrotropy is not exactly zero for regions far away of the reconnection with a background noise around $0.1$, while the detection algorithm clearly identifies single distributions.
\begin{figure} [ht]   
\gridline{
  \includegraphics[trim={0.5cm 3.7cm 1.25cm 3.85cm},clip,width=0.5\linewidth]{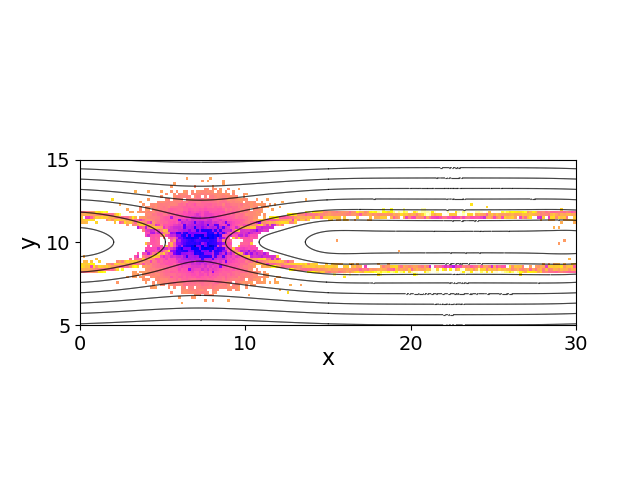}
  \includegraphics[trim={0.5cm 3.7cm 1.25cm 3.85cm},clip,width=0.5\linewidth]{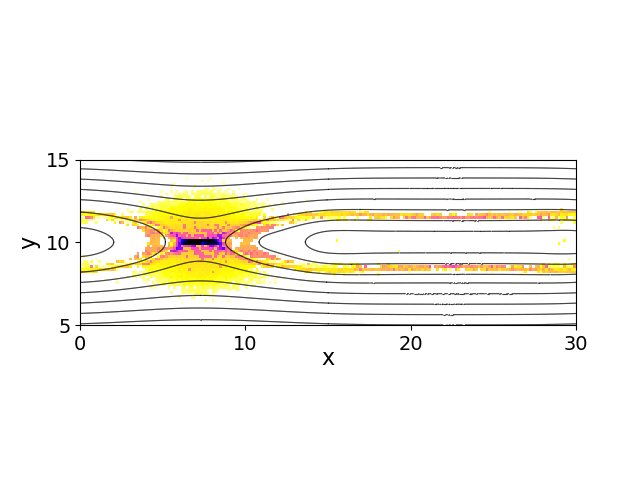}
}

\vspace{-0.35cm}

\gridline{
  \includegraphics[trim={0.5cm 3.7cm 1.25cm 3.85cm},clip,width=0.5\linewidth]{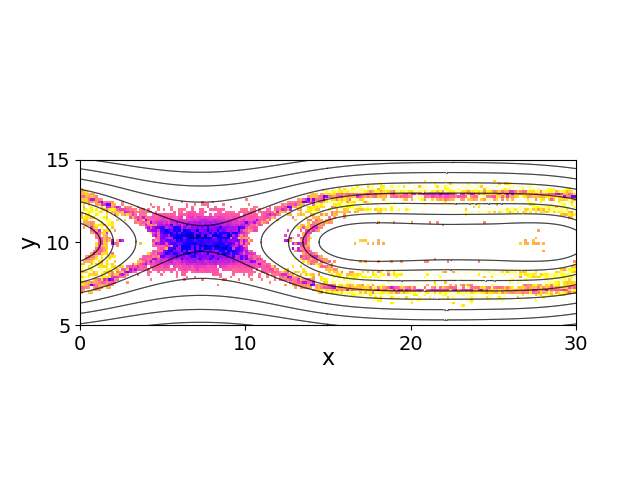}
  \includegraphics[trim={0.5cm 3.7cm 1.25cm 3.85cm},clip,width=0.5\linewidth]{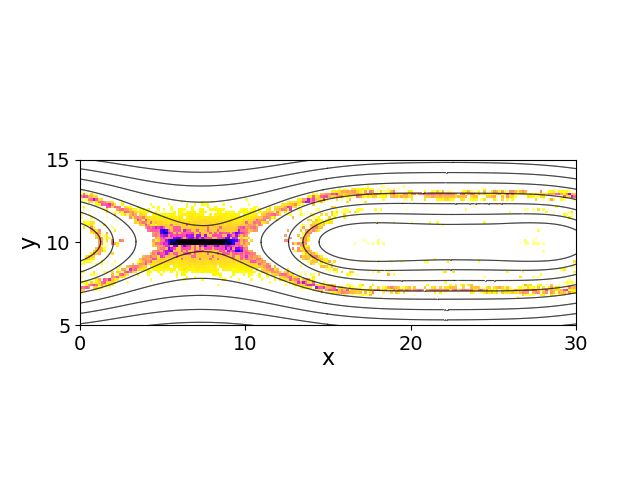}
}

\vspace{-0.35cm}

\gridline{
  \includegraphics[trim={0.5cm 3.7cm 1.25cm 3.85cm},clip,width=0.5\linewidth]{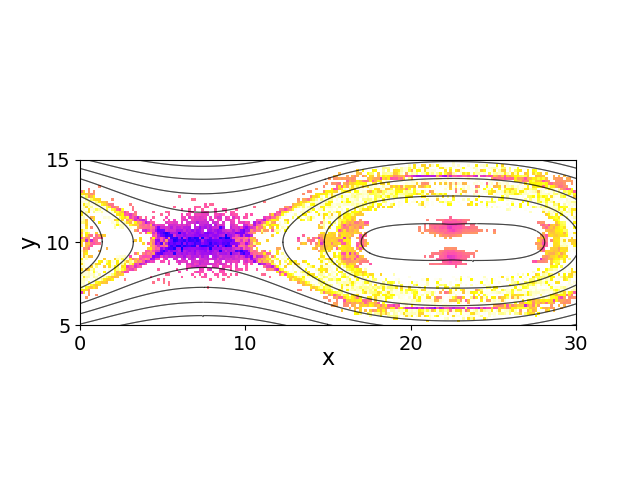}
  \includegraphics[trim={0.5cm 3.7cm 1.25cm 3.85cm},clip,width=0.5\linewidth]{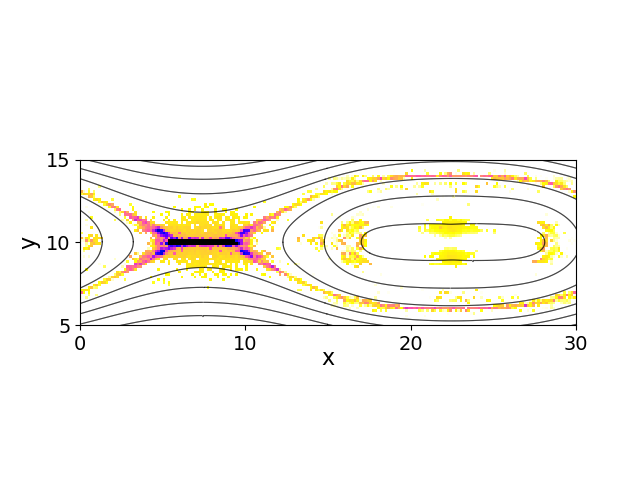}
}

\vspace{-0.35cm}

\gridline{
  \includegraphics[trim={0.5cm 1.5cm 1.25cm 3.65cm},clip,width=0.5\linewidth]{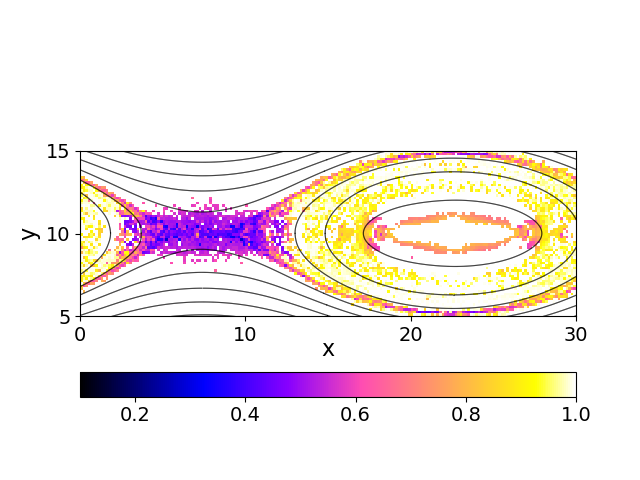}
  \includegraphics[trim={0.5cm 1.5cm 1.25cm 3.65cm},clip,width=0.5\linewidth]{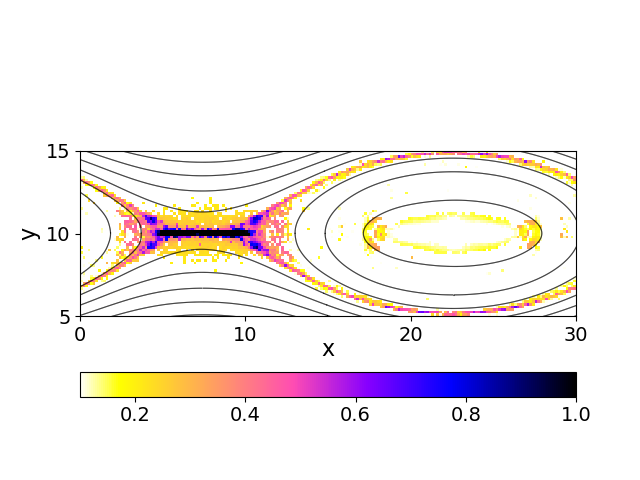}
}    
\caption{The left-hand column highlights the energy drop $E_{drop}$ defined by~\autoref{eq:energy_drop} and the right-hand column depicts the energy deviation $E_{dev}$ given by~\autoref{eq:energy_deviation}. Both quantities are presented at four different time steps, from top to bottom: $t=8,000$, $t=12,000$, $t=16,000$, and $t=20,000$.}
\label{fig:dh_result_weak_thermal} 
\end{figure} 

\autoref{fig:dh_result_weak_thermal} displays $E_{drop}$ and $E_{dev}$ in order to support the analysis of the number of components, helping to make distinctions between the different distributions. First, the result for $E_{drop}$ on the left-hand column maps a large region around the ion diffusion region and the EDR with low values around 0.5, reflecting potential particle accelerations. Moreover, a very narrow region matching the EDR definition of the gyrotropy is spotted by high values of $E_{dev}$. Maximum values are truncated by the color bar, but they exceed 3.0. Thus, the algorithm clearly spots a specific region related to the EDR, similar to the measure of gyrotropy results, and also differentiates the ion diffusion region from the EDR with very different $E_{dev}$ values. In particular, distributions in the EDR are expected to be very complex with mixtures distributed all over the velocity space, while clustered mixtures can be predicted for the ion diffusion region. For the two regions, mixtures with a small extend relatively to the second statistical moment of the overall distribution are expected. 

Energy deviation and energy drop provide also valuable informations for distributions around the O point. In particular, the very low $E_{dev}$ values (around $0.0$) in this region suggest that a significant number of distributions tagged with $2$ and $3$ components should not be interpreted as beams but rather as a single distribution deviated from a Maxwellian. The high energy drop ratio (about $0.9$) supports this assumption as these clustered distributions show a thermal energy very close to the thermal energy of a Maxwellian. As regards the outflow, quite small $E_{drop}$ values about $0.6$ are observed coupled with significant $E_{dev}$ around $0.8$. We can therefore assume that the mixtures have a pattern similar to the EDR but to a lesser extent. Finally, $E_{drop}$ ranges between $0.8$ and $1.0$ for the separatrix region while $E_{dev}$ shows an upper bound of about $0.4$, half as much as the outflow. Thus, distributions in this region seem to tend to a single population but still being very different from a Maxwellian.

\begin{figure} [ht]   
\gridline{
  \includegraphics[trim={4.3cm 1.63cm 5.9cm 0.35cm},clip,width=0.283\linewidth]{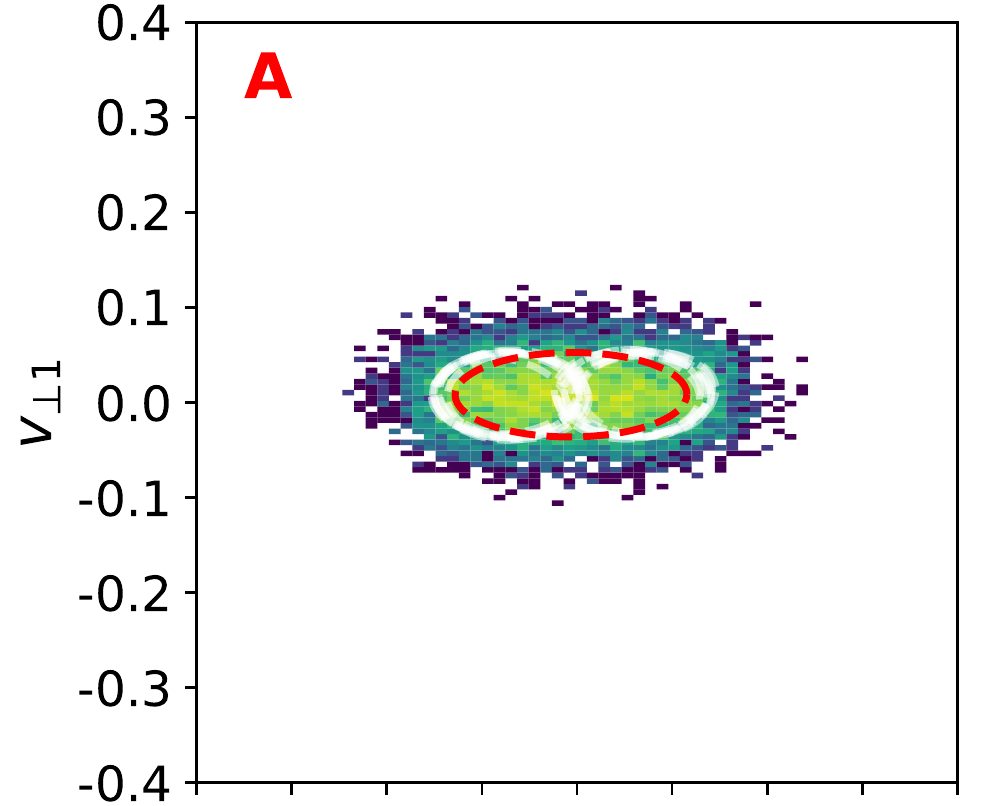}
 \includegraphics[trim={4.3cm 1.63cm 5.9cm 0.35cm},clip,width=0.283\linewidth]{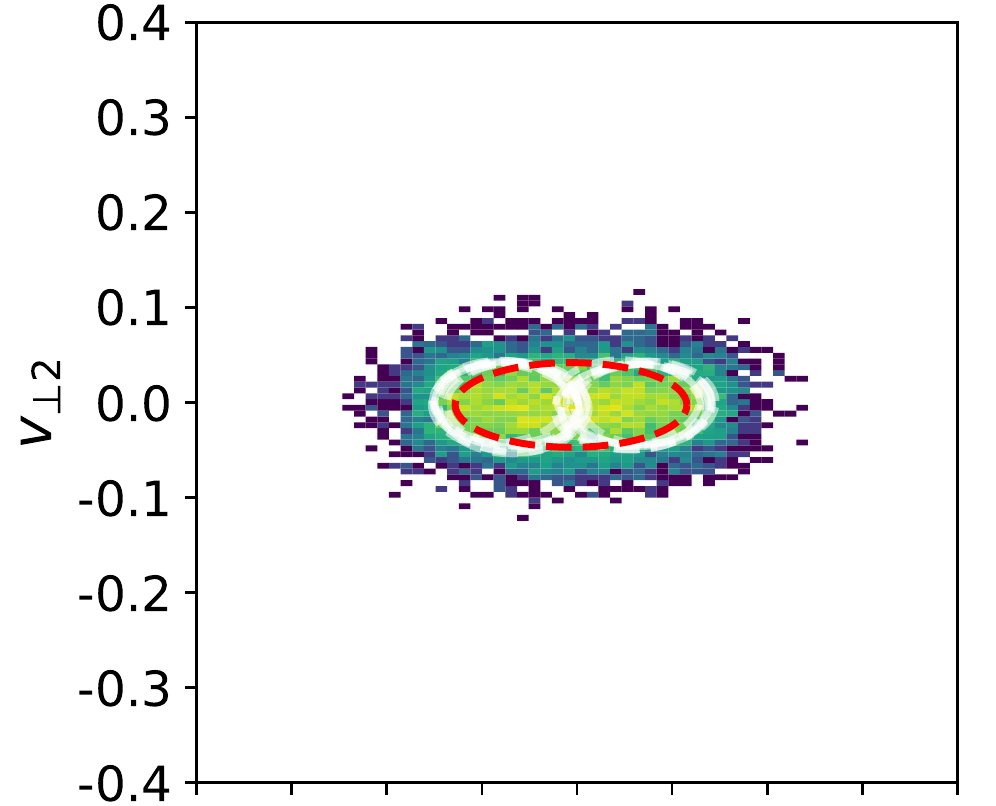}
  \includegraphics[trim={4.1cm 1.63cm 4.9cm 0.35cm},clip,width=0.316\linewidth]{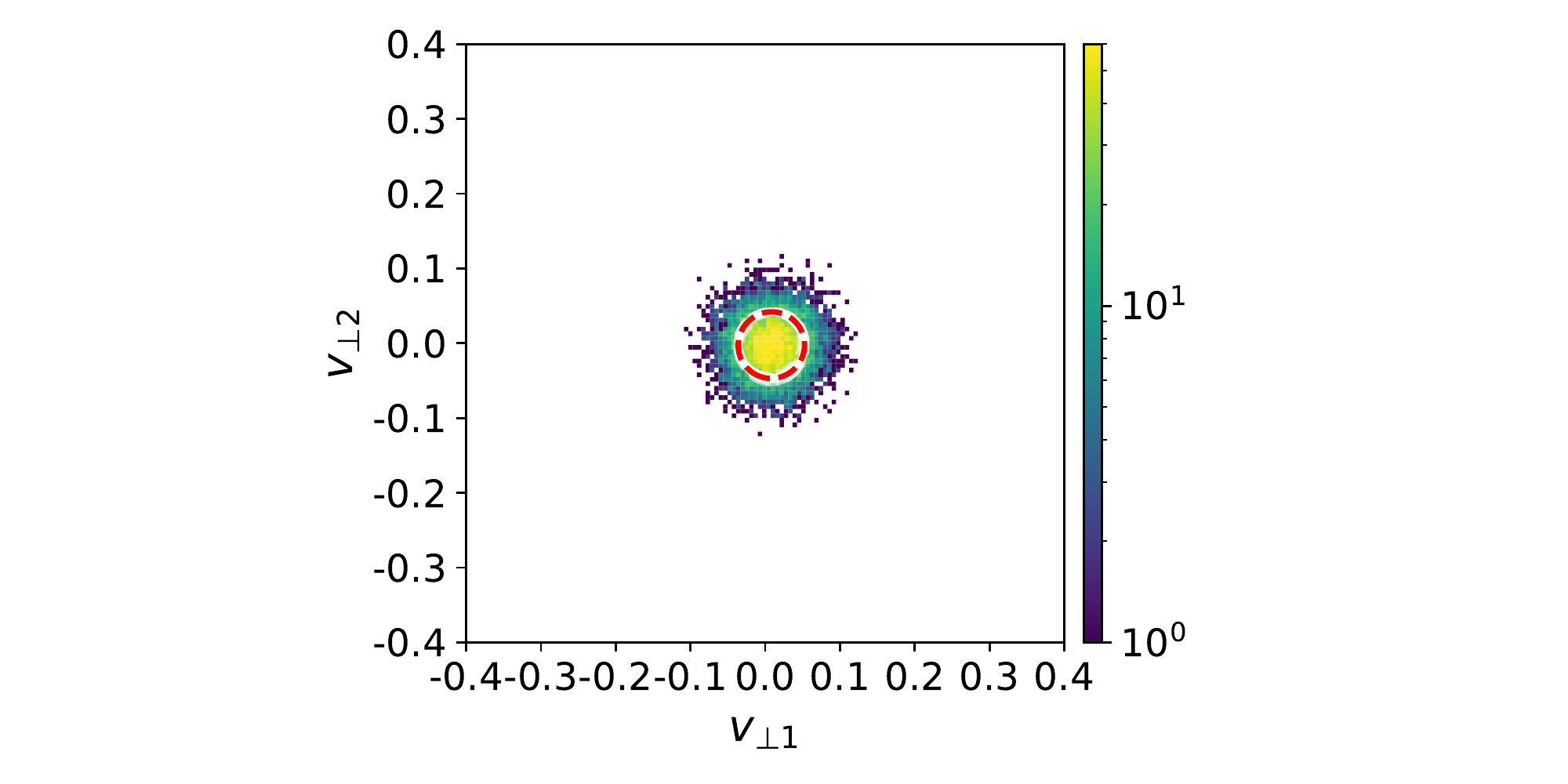}
} 

\vspace{-0.3cm}

\gridline{
 \includegraphics[trim={4.3cm 1.63cm 5.9cm 0.35cm},clip,width=0.283\linewidth]{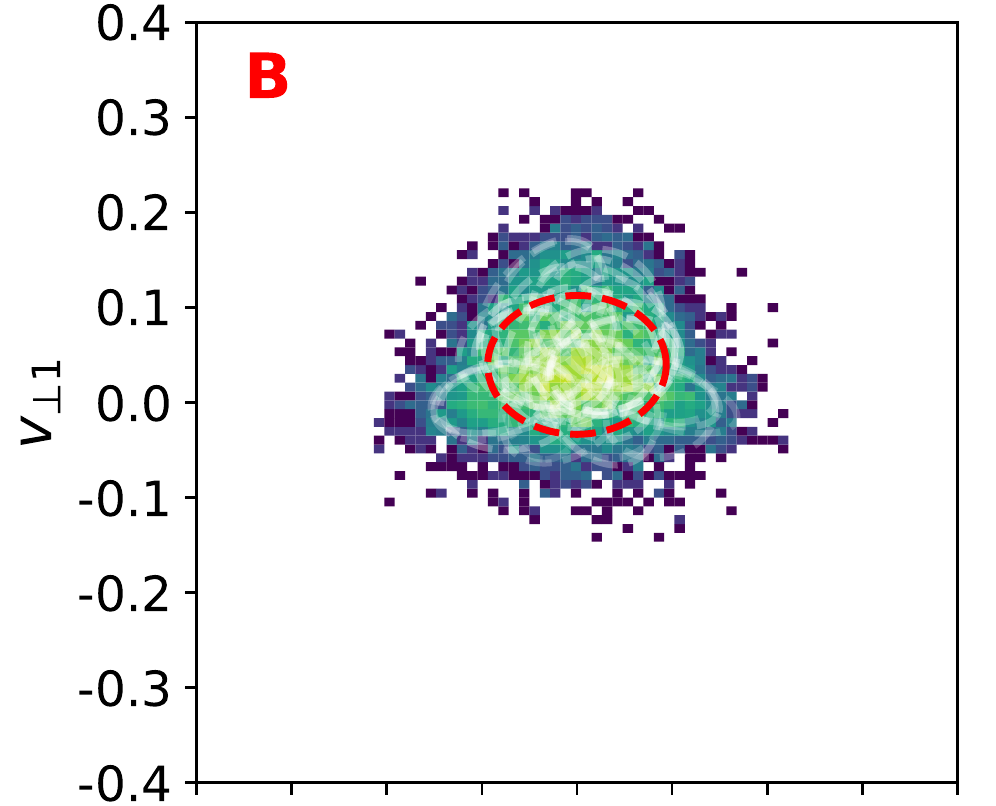}
 \includegraphics[trim={4.3cm 1.63cm 5.9cm 0.35cm},clip,width=0.283\linewidth]{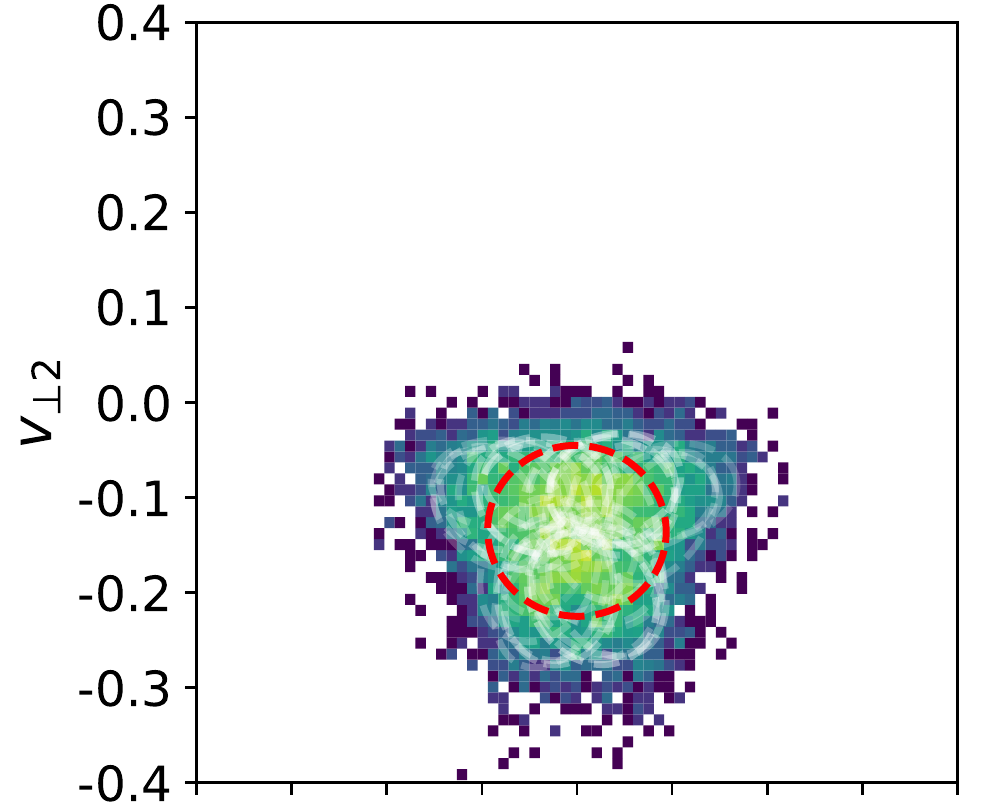}
 \includegraphics[trim={4.1cm 1.63cm 4.9cm 0.35cm},clip,width=0.316\linewidth]{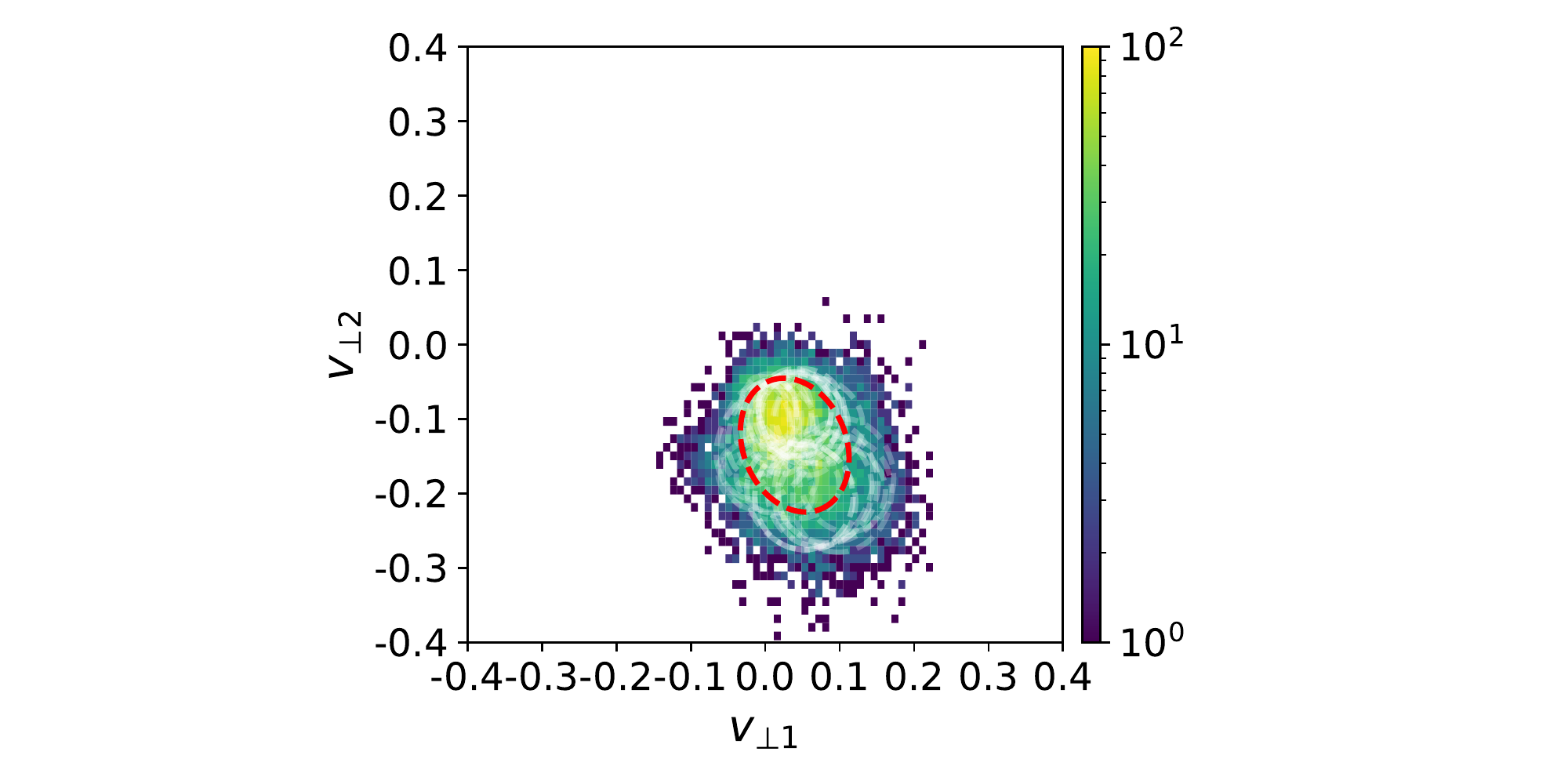}
}

\vspace{-0.3cm}

\gridline{
 \includegraphics[trim={4.3cm 1.63cm 5.9cm 0.35cm},clip,width=0.283\linewidth]{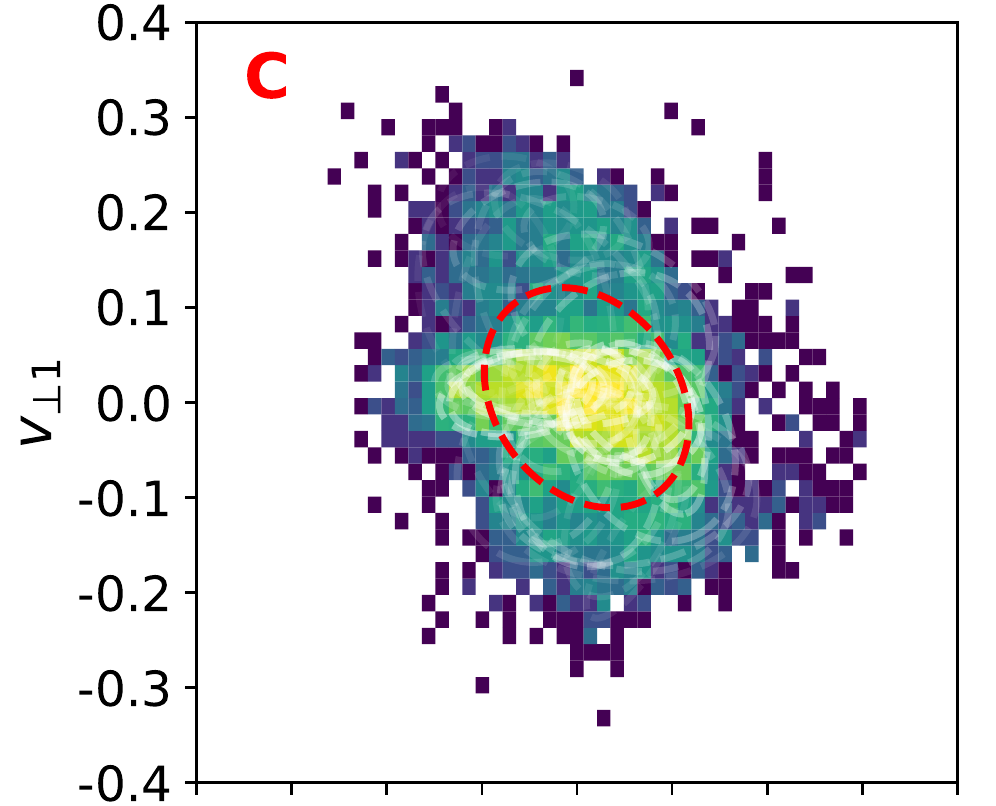}
 \includegraphics[trim={4.3cm 1.63cm 5.9cm 0.35cm},clip,width=0.283\linewidth]{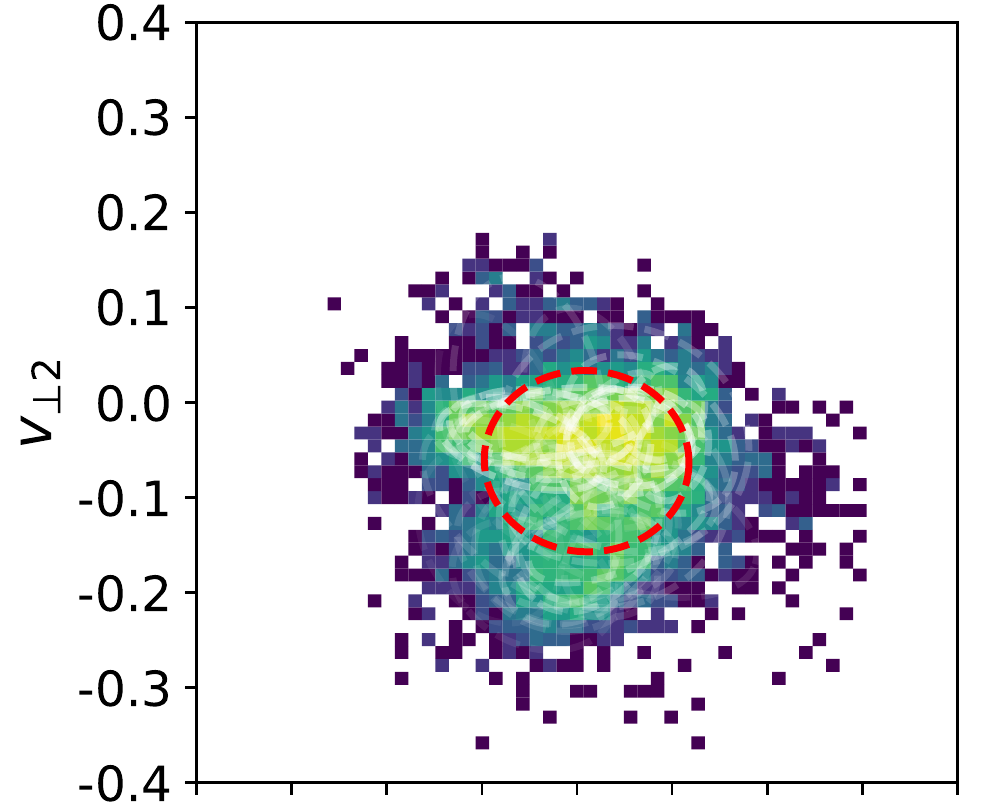}
  \includegraphics[trim={4.1cm 1.63cm 4.9cm 0.35cm},clip,width=0.316\linewidth]{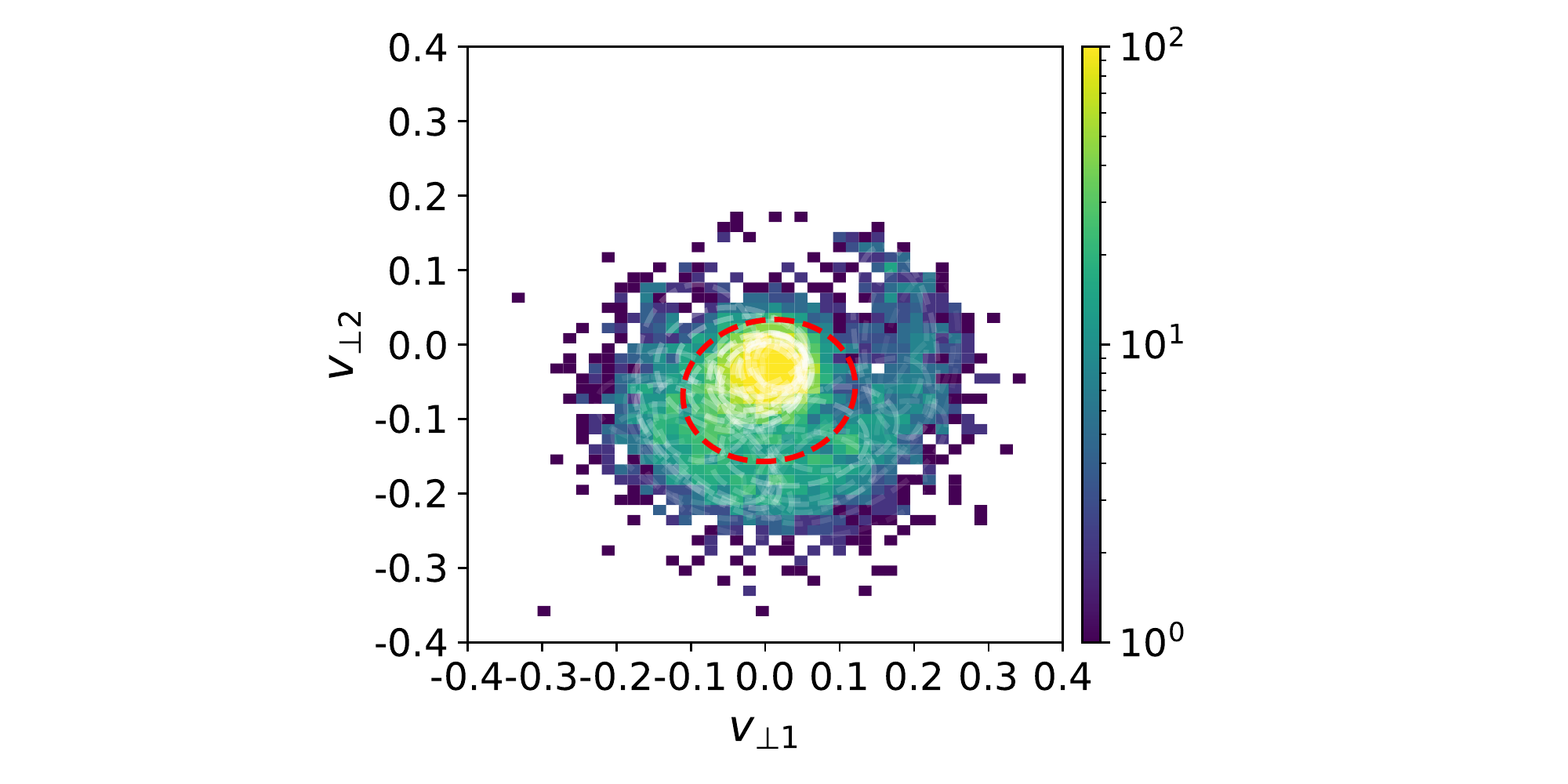}
}

\vspace{-0.3cm}

\gridline{
 \includegraphics[trim={4.3cm 1.63cm 5.9cm 0.35cm},clip,width=0.283\linewidth]{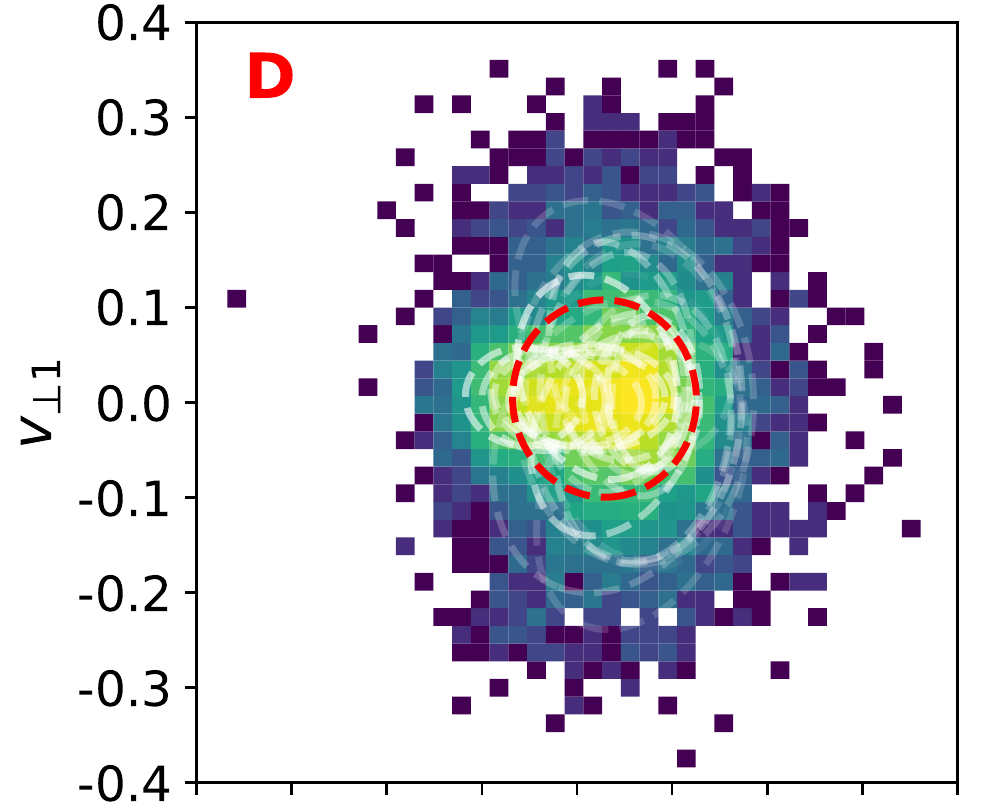}
 \includegraphics[trim={4.3cm 1.63cm 5.9cm 0.35cm},clip,width=0.283\linewidth]{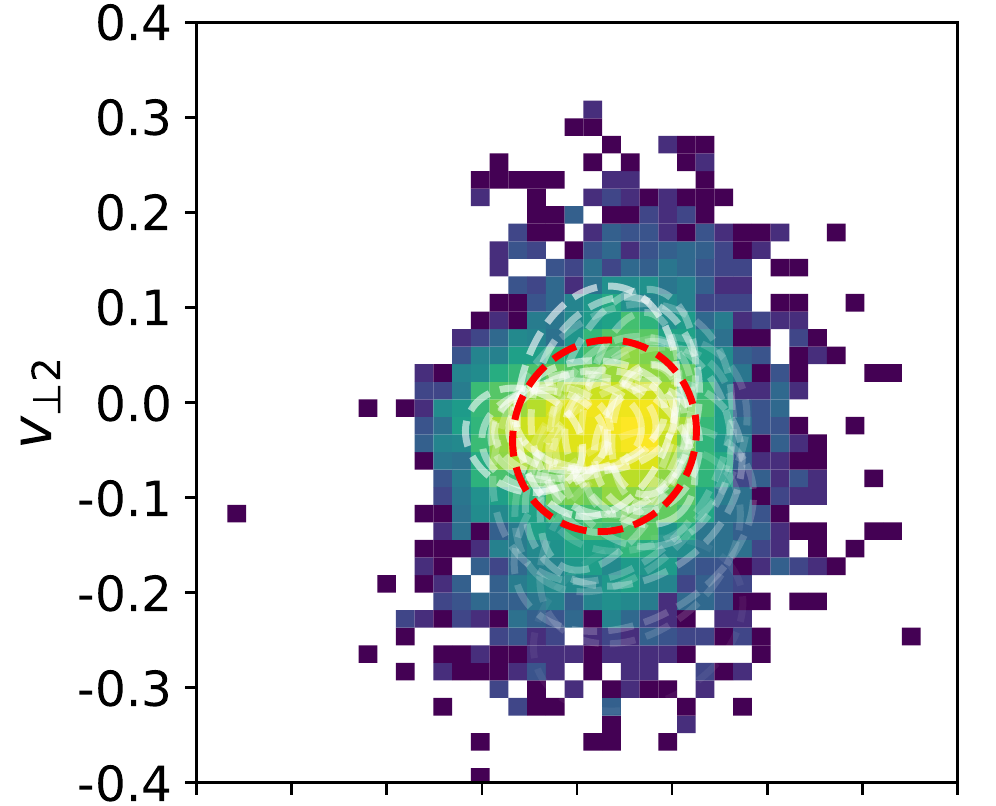}
   \includegraphics[trim={4.1cm 1.63cm 4.9cm 0.35cm},clip,width=0.316\linewidth]{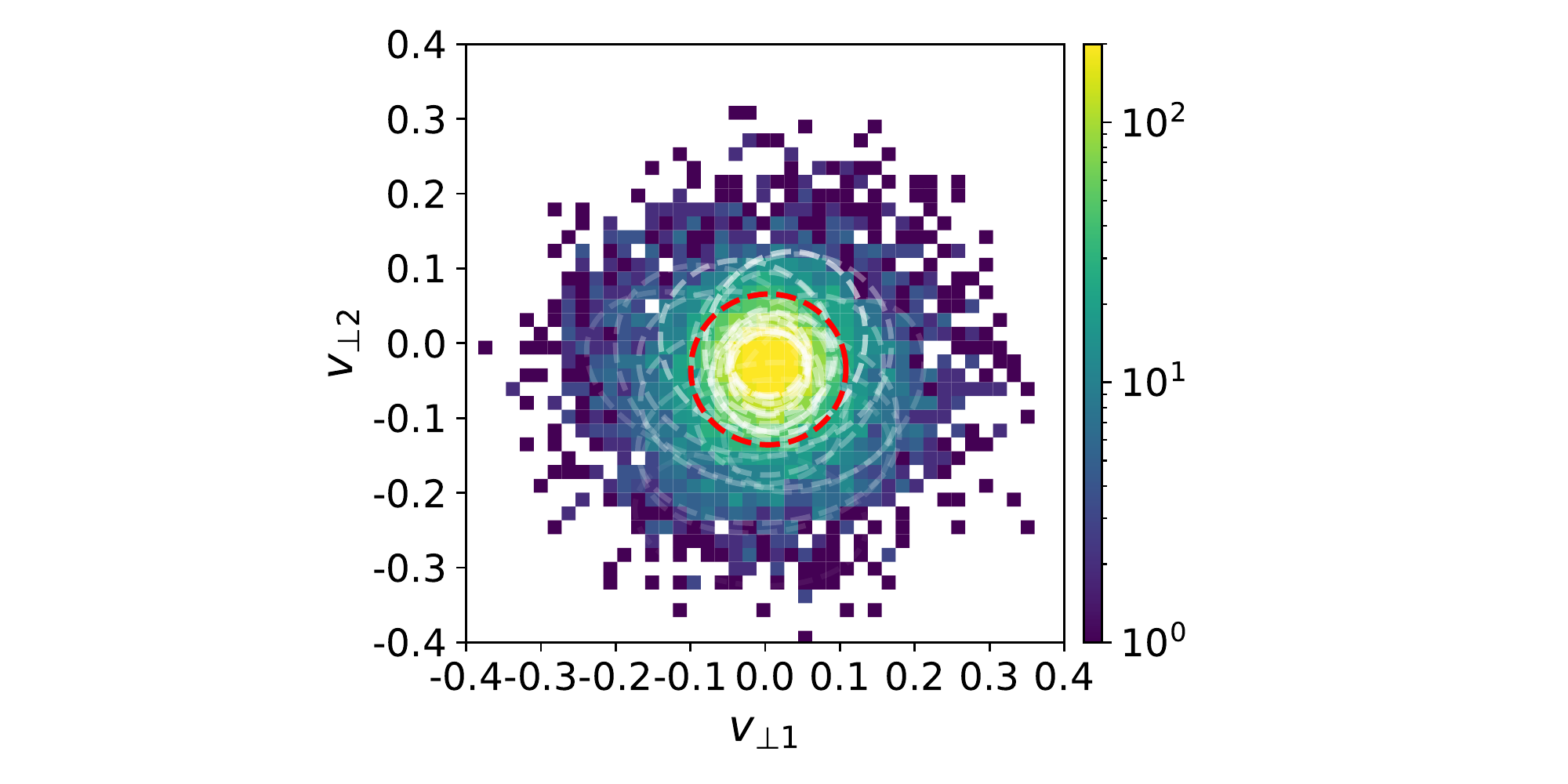}
}

\vspace{-0.3cm}

\gridline{
 \includegraphics[trim={4.3cm 0.47cm 5.9cm 0.35cm},clip,width=0.283\linewidth]{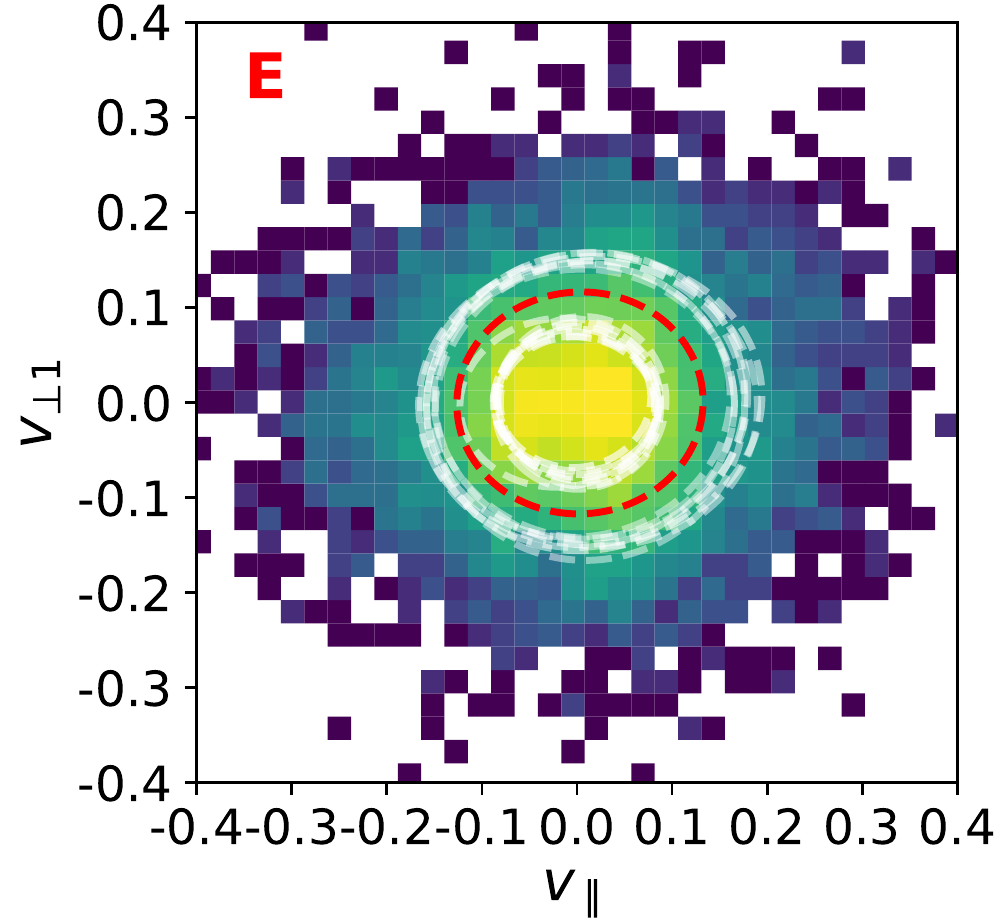}
 \includegraphics[trim={4.3cm 0.47cm 5.9cm 0.35cm},clip,width=0.283\linewidth]{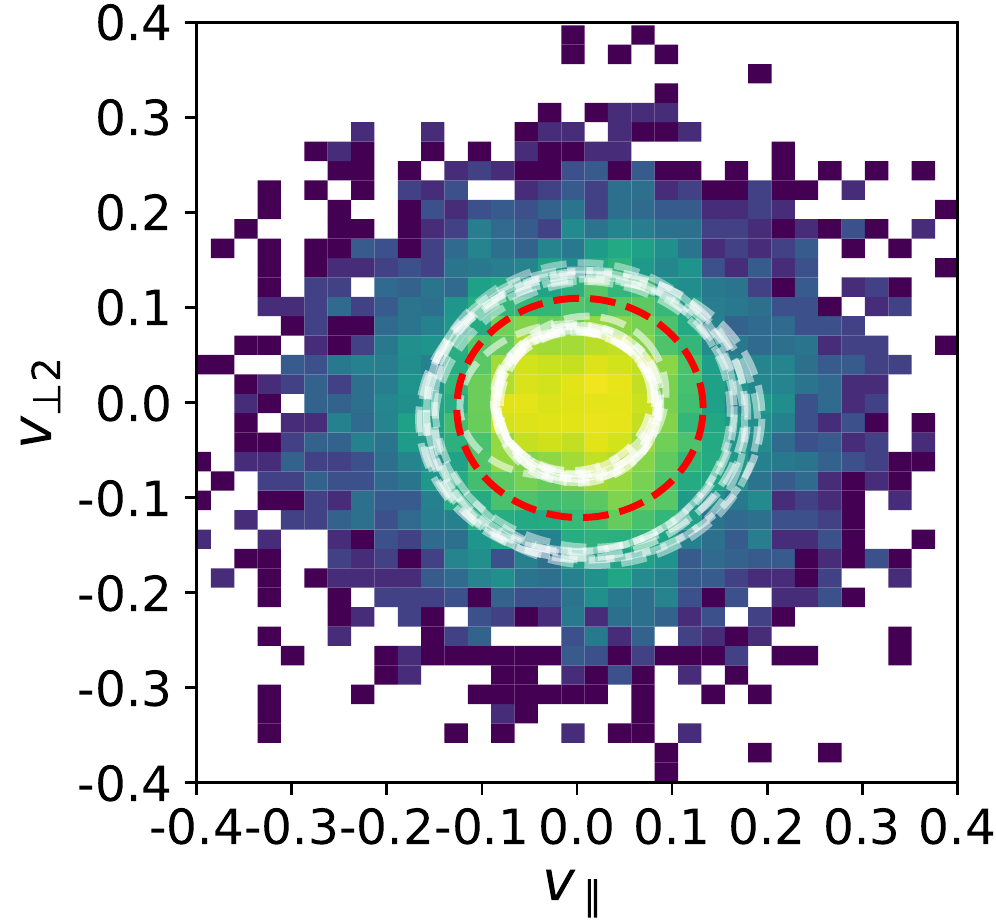}
  \includegraphics[trim={4.1cm 0.47cm 4.9cm 0.35cm},clip,width=0.316\linewidth]{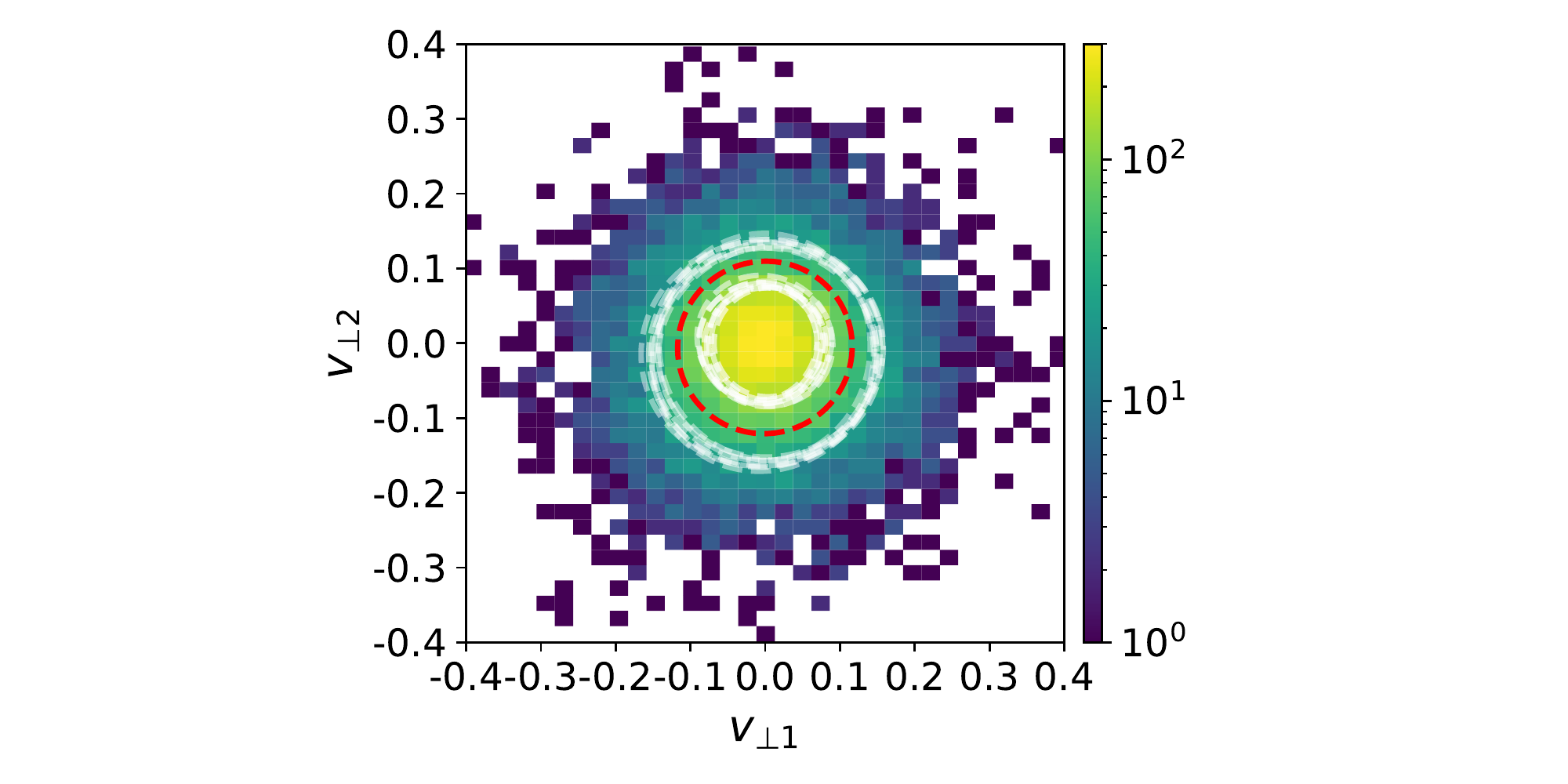}
}
\caption{Electron velocity distribution for the Double Harris sheet case at $t=20,000$. Each row correspond to one of the five red rectangles depicted in~\autoref{fig:dh_result_weak}. Three two dimensional marginal distributions are presented:  $v_{\parallel}-v_{\bot 1}$, $v_{\parallel}-v_{\bot 2}$, and $v_{\bot 1}-v_{\bot 2}$. The white ellipses illustrate the different Gaussians of the mixtures in each distribution. The transparency is determined by the weight of each Gaussian: no transparency for a weight of $1$ and a full transparency for a zero weight. The red ellipses give the mean and variance for a single distribution.}
\label{fig:dh_distri_week} 
\end{figure}

\autoref{fig:dh_distri_week} illustrates the different distributions associated to the five red rectangles displayed in~\autoref{fig:dh_result_weak}. These rectangles depict distributions by considering $4$ square windows of the algorithm in the $x$ direction and $2$ square windows in the $y$ direction. Several observations can be made:
\begin{itemize}
    \item [-] \textbf{Inflow region} (box 'A'): the distributions show a strong anisotropy demonstrating the heating along the parallel direction $v_{\parallel}$. This behavior is characteristic for the inflow region~\citep{egedal2016spacecraft}. However, the GMM algorithm does not explain the data with a unique anisotropic Gaussian but uses two components to fit the broad mode with the short tail. It may suggest the electron distribution already deviates from a Maxwellian within the inflow region. The short tail approximated by two components leads to high $E_{drop}$ values as each mixture is relatively small compared to the overall second moment. The other directions perpendicular to the magnetic field show a Gaussian shape.
    \item [-] \textbf{Electron distribution region} (box 'B'): A crescent shape seems to be observed in the $v_{\bot 1}-v_{\bot 2}$ marginal distribution, justifying the high number of components identified by the algorithm. As expected from the $E_{drop}$ and $E_{dev}$ analysis, all the mixtures are spread over the distribution with a small width compared to the second moment computed over the whole distribution which cannot properly fit such a complex distribution. Triangular shapes are depicted in the two other projections $v_{\parallel}-v_{\bot 1}$ and $v_{\parallel}-v_{\bot 2}$.
    
    \item [-] \textbf{Outflow region} (box 'C'): A crescent shape is also observed justifying why this region has a number of components similar to the EDR. A narrow Gaussian associated to a high weight fits the zero-centered mode of the distribution in the $v_{\bot 1}-v_{\bot 2}$ projection while scattered Gaussians with low weights are dedicated to the crescent shape. The two other projections are quite complex due to the crescent-shape. From the algorithm point of view, the crescent-shape is hard to approximate, that is why 4 components are needed: one of the core and the three other for the crescent.
    
    \item [-] \textbf{First separatrix region} (box 'D'): The crescent shape has disappeared in the separatrix region. However, complex distributions are observed in the $v_{\parallel}-v_{\bot 1}$ and $v_{\parallel}-v_{\bot 2}$ projections: the parallel direction has a large peak while elongated tails are observed in the two other directions. The three components are particularly needed to fit the large peak.
    
    \item [-] \textbf{Second separatrix region} (box 'E'): The distribution deviates only slightly from a Maxwellian but the mixture acts the opposite of the inflow. Thin core distributions associated with strong weights are supplemented by wider distributions with smaller weights. This pair of distribution (core and tail) improves the overall fitting, in particular the long tail. This observation validates the results from $E_{drop}$ and $E_{dev}$: the mixture of the GMM variances is very close to the second order while the different components are almost all zero-centered, leading to a very low $E_{dev}$ value.
\end{itemize}


\section{Conclusion}
We have proposed to automatically identify magnetic reconnection from velocity particle distributions using a density estimation technique called Gaussian Mixture Model. This approach has been able to identify various different regions around reconnection sites provided by a PIC simulation with a weak guide field, but also for a strong guide field. Analyzing the thermal heating and the distributions of the different components of the Gaussian mixtures gives a physical interpretation beyond the pure statistical properties of the GMM, helping to distinguish between heating and accelerating particles into beams but also between unimodal distributions and complex distributions. This method represents one of the first application of machine learning algorithms to particle distributions in plasma physics. Nevertheless, it does not represent a unique solution to detect unambiguously magnetic reconnection. A central part of the algorithm is based on the Bayesian Information Criterion which is sensitive to the number of particles, 
Therefore, the algorithm may require a calibration to properly set the resolution with regards to the number of particles and available data.  

For the moment, only 2.5D simulations have been investigated, but there is no reason to doubt that the approach cannot be applied to other types of simulation. Thus, testing the algorithm with three-dimensional simulations represent a mandatory next step. Other fields of application could also be proposed, such as turbulence analysis. Moreover, simulations have access to the complete description of the plasma over all the spatial grid, while in-situ observations are restricted to a a small set of measurements at specific spacecraft locations. Therefore, further works will also focus on the application of this detection algorithm to local observational data of particle distribution function. 




\acknowledgments
\section*{Acknowledgments}

This paper has received funding from the European Union’s Horizon 2020 research and innovation programme under grant agreement No 776262 (AIDA, www.aida-space.eu). This research used resources of the National Energy Research Scientific Computing Center, which is supported by the Office of Science of the US Department of Energy under Contract no. DE-AC02-05CH11231. Additional computing has been provided by NASA NAS and NCCS High Performance Computing, by the Flemish Supercomputing Center (VSC), and by PRACE Tier-0 allocations.


\appendix

\section{Strong guide field}
\label{app:strong_guide}
\begin{figure} [ht]   
\gridline{
  \includegraphics[trim={0.55cm 3.75cm 1.25cm 3.85cm},clip,width=0.5\linewidth]{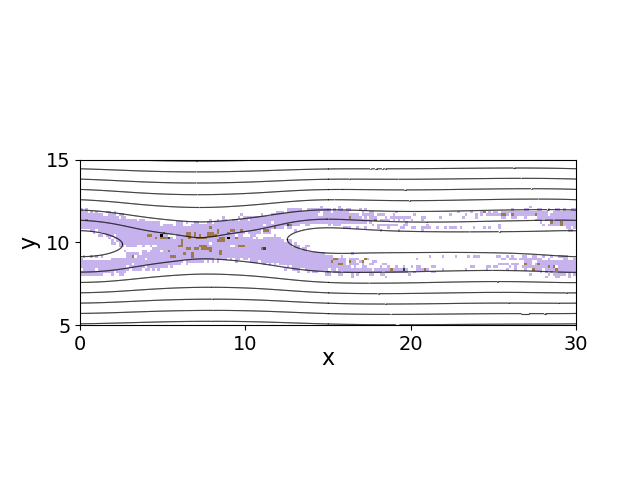}
    \includegraphics[trim={0.55cm 3.75cm 1.25cm 3.85cm},clip,width=0.5\linewidth]{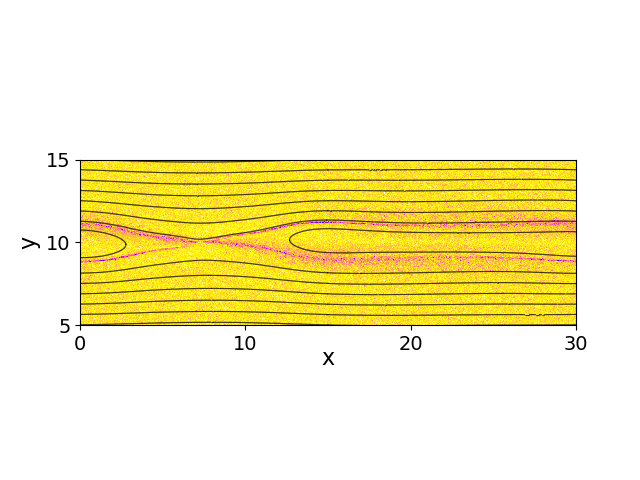}
}

\vspace{-0.35cm}

\gridline{
  \includegraphics[trim={0.55cm 3.75cm 1.25cm 3.85cm},clip,width=0.5\linewidth]{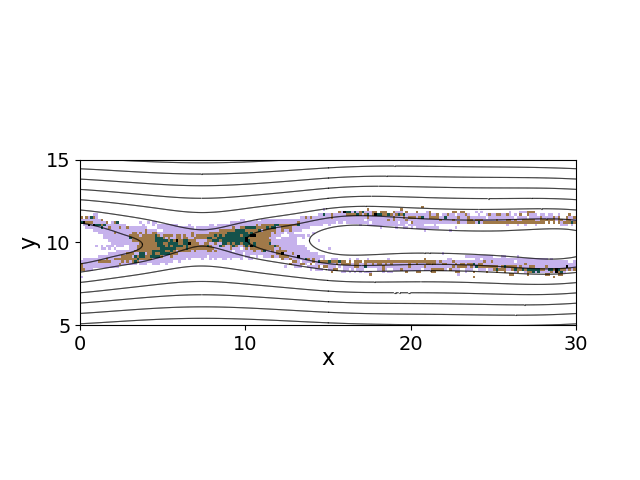}
   \includegraphics[trim={0.55cm 3.75cm 1.25cm 3.85cm},clip,width=0.5\linewidth]{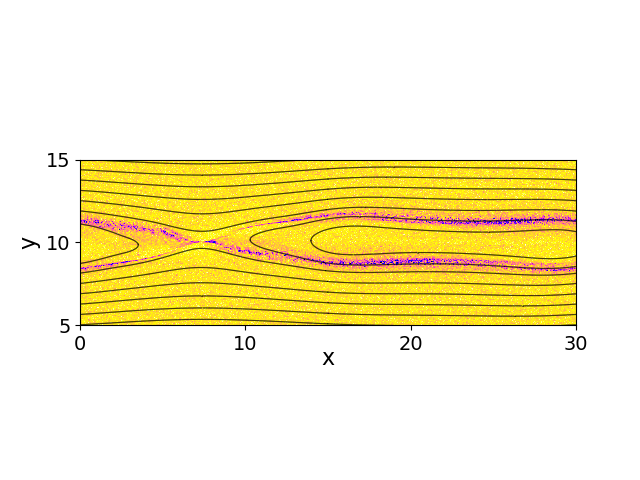}
}

\vspace{-0.35cm}

\gridline{
  \includegraphics[trim={0.55cm 3.75cm 1.25cm 3.85cm},clip,width=0.5\linewidth]{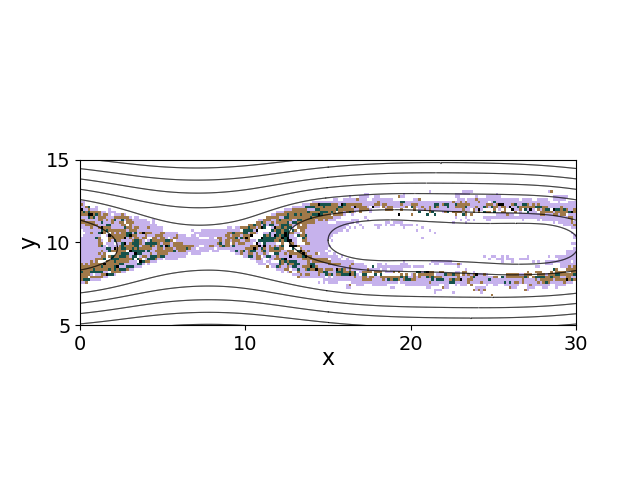}
    \includegraphics[trim={0.55cm 3.75cm 1.25cm 3.85cm},clip,width=0.5\linewidth]{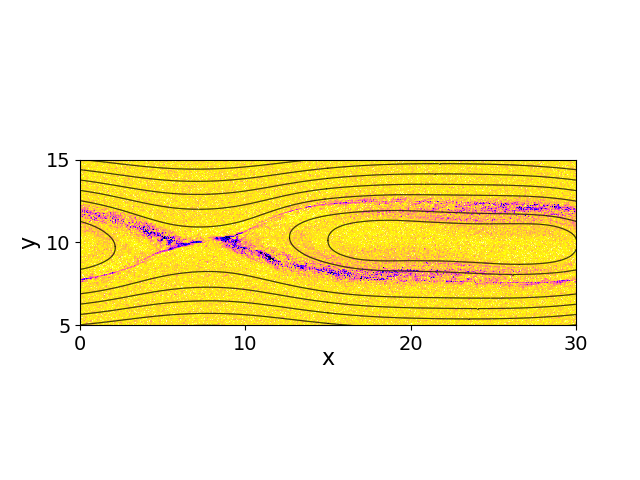}
}

\vspace{-0.35cm}

\gridline{
  \includegraphics[trim={0.55cm 1.5cm 1.25cm 3.65cm},clip,width=0.5\linewidth]{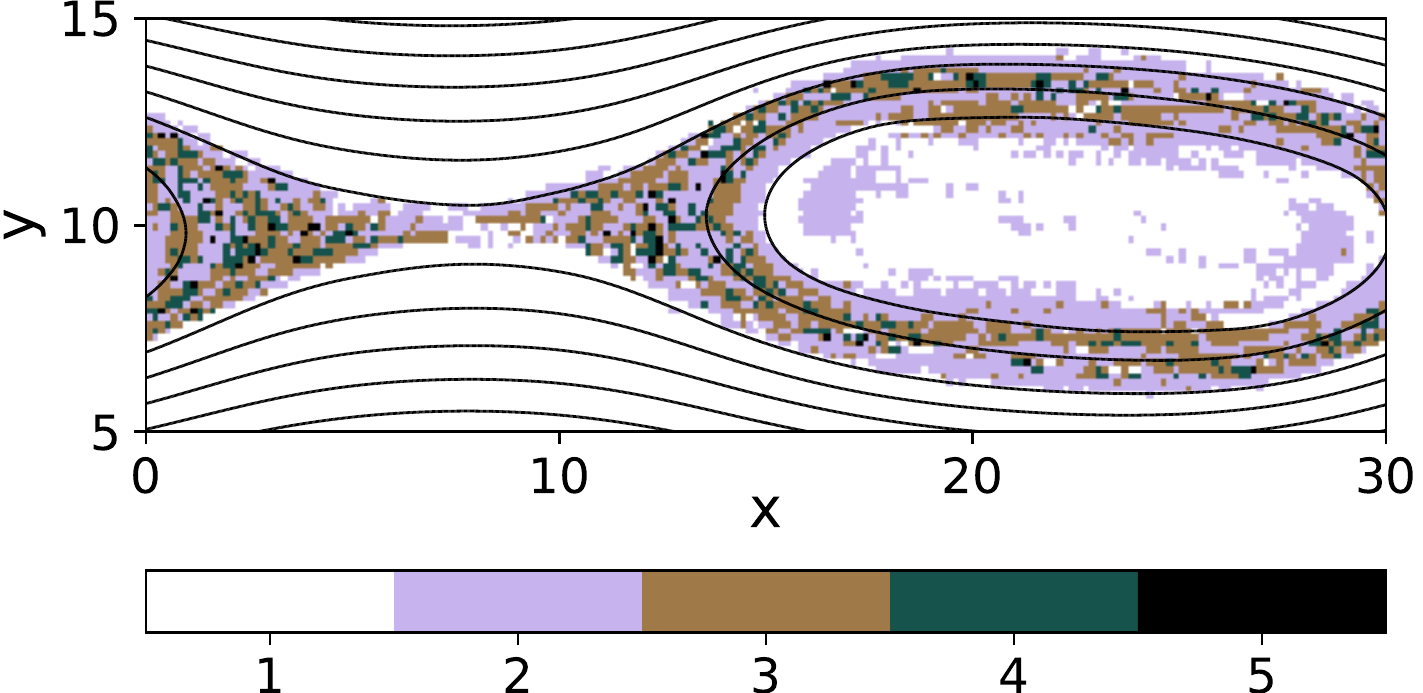}
    \includegraphics[trim={0.55cm 1.5cm 1.25cm 3.65cm},clip,width=0.5\linewidth]{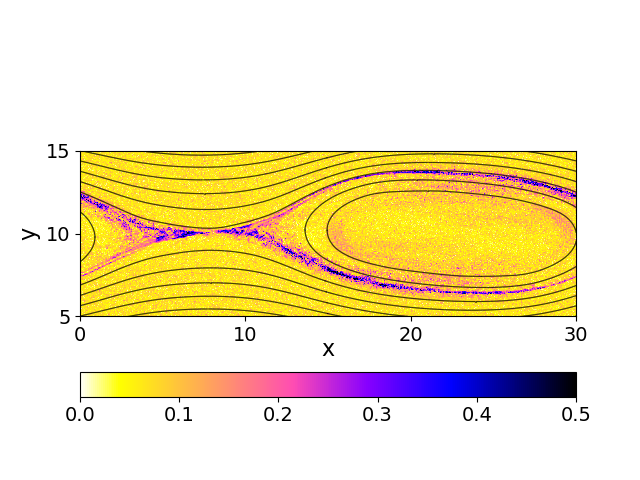}
}
\setlength{\belowcaptionskip}{-8pt}
 \caption{Magnetic reconnection detection for the Double Harris sheet case with a strong guide field at four different time steps, from top to bottom: $t=8,000$, $t=12,000$, $t=16,000$, and $t=20,000$. The left-hand column presents the number of components provided by the BIC optimization and the right-hand column shows the measure of gyrotropy $\sqrt{Q}$ defined in~\autoref{eq:mesure_gyro}. The red rectangles indicates the location where specific distributions are observed.}
\label{fig:app_dh_result_strong}    
\end{figure}

As a second case we consider the same Double Harris Sheet simulation but with a strong guide field.~\autoref{fig:app_dh_result_strong} presents the same quantities than~\autoref{fig:dh_result_weak} but for the strong guide field: the number of components on the left-hand column and the measure of gyrotropy on the right-hand column. The regions identified by the algorithm are very different. The inflow has almost vanished while the EDR is very hard to identify. The outflow seems to be present, in particular at $t=12,000$ but even this region tends to disappear for the other time steps, leading to a quite noisy distribution of the number of components. As in the previous weak guide field case, the measure of gyrotropy shares roughly the same topological boundaries, although the detection algorithm identifies wider regions, in particular around the EDR for the small time steps at $t=8,000$ and $t=12,000$.

\begin{figure} [ht]   
\gridline{
  \includegraphics[trim={0.5cm 3.7cm 1.25cm 3.85cm},clip,width=0.5\linewidth]{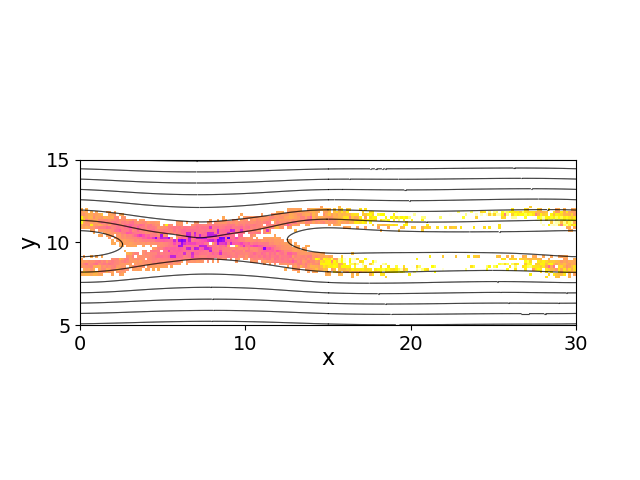}
  \includegraphics[trim={0.5cm 3.7cm 1.25cm 3.85cm},clip,width=0.5\linewidth]{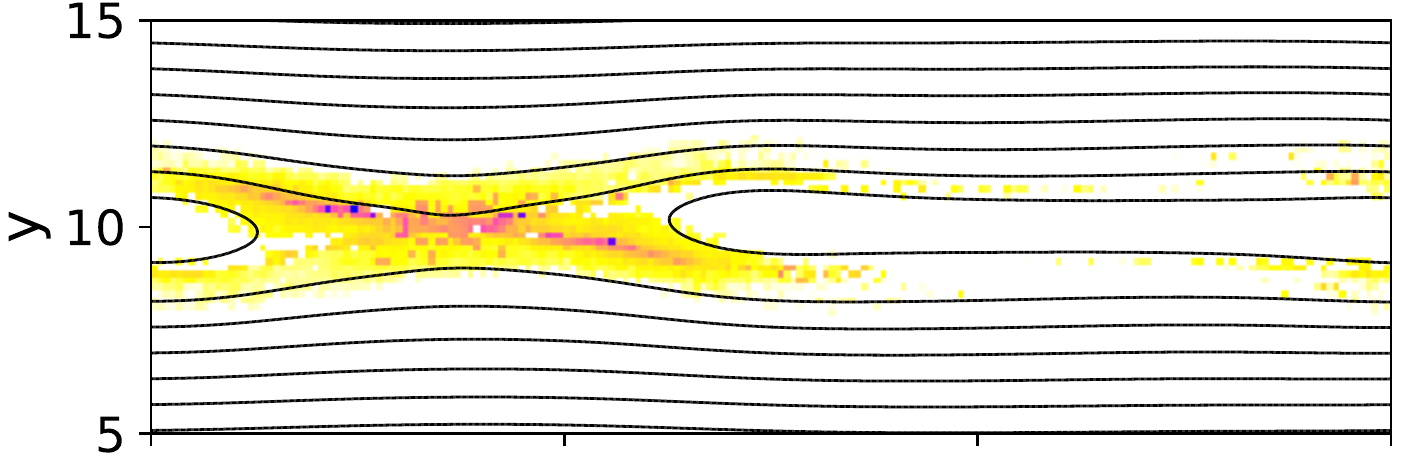}
}

\vspace{-0.35cm}

\gridline{
  \includegraphics[trim={0.5cm 3.7cm 1.25cm 3.85cm},clip,width=0.5\linewidth]{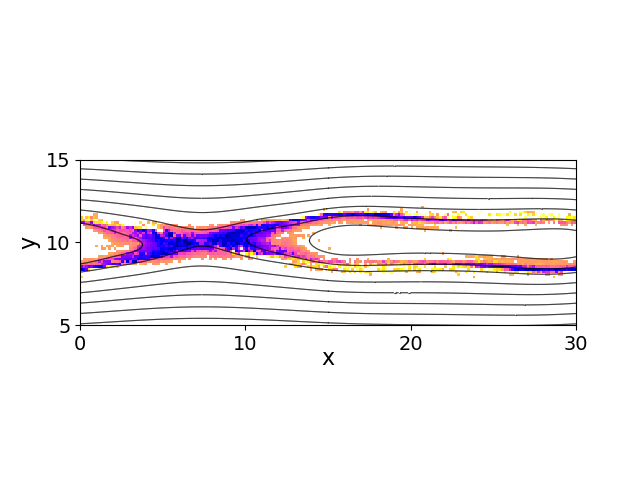}
  \includegraphics[trim={0.5cm 3.7cm 1.25cm 3.85cm},clip,width=0.5\linewidth]{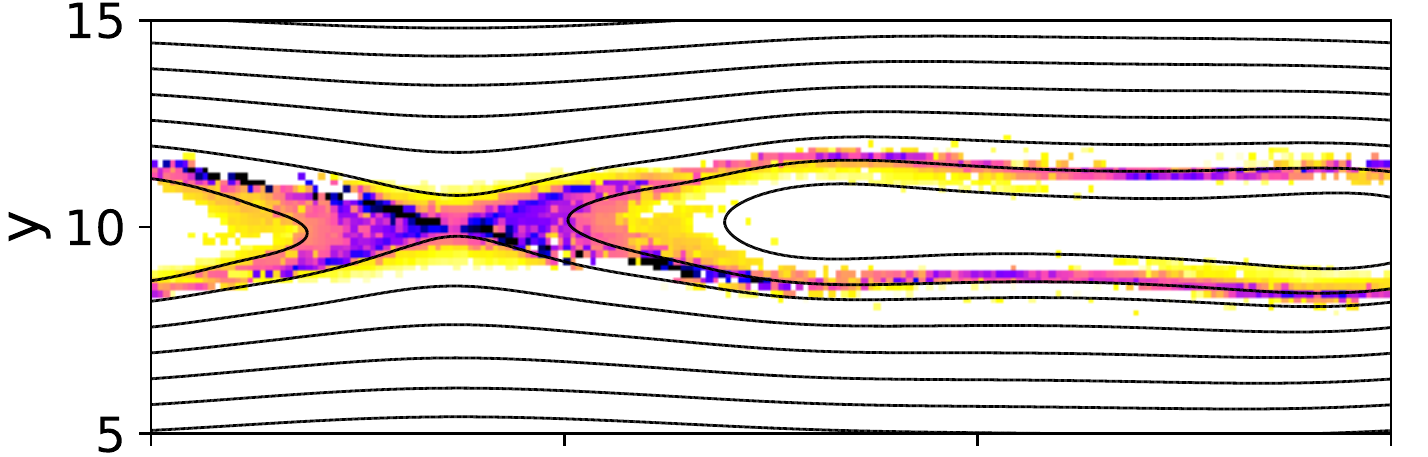}
}

\vspace{-0.35cm}

\gridline{
  \includegraphics[trim={0.5cm 3.7cm 1.25cm 3.85cm},clip,width=0.5\linewidth]{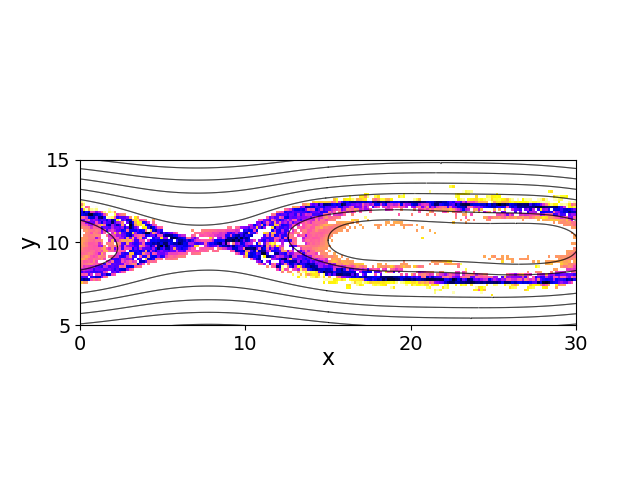}
  \includegraphics[trim={0.5cm 3.7cm 1.25cm 3.85cm},clip,width=0.5\linewidth]{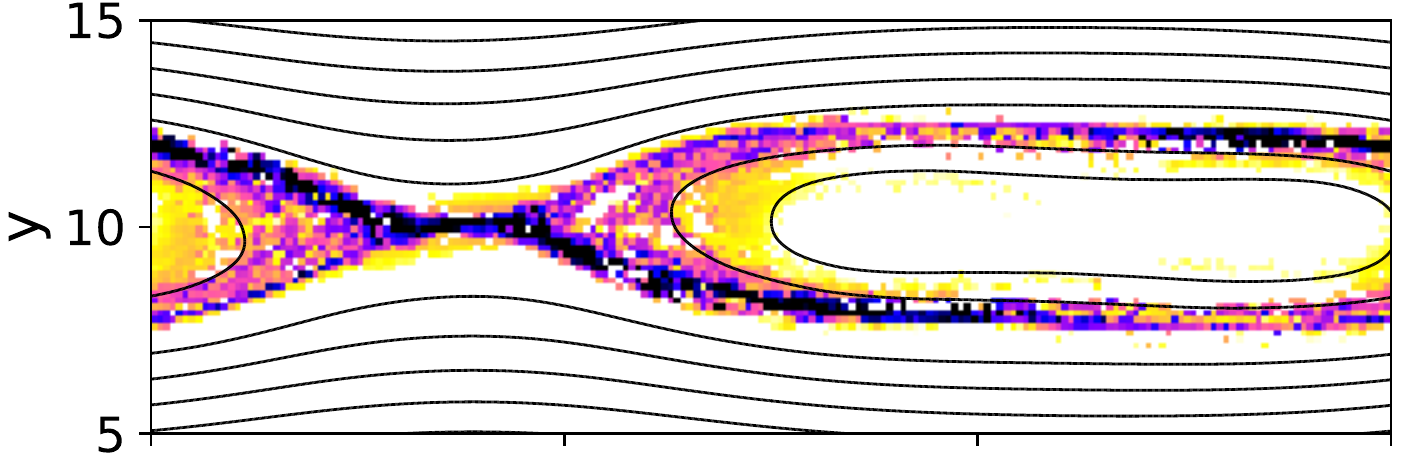}
}

\vspace{-0.35cm}

\gridline{
  \includegraphics[trim={0.5cm 1.5cm 1.25cm 3.65cm},clip,width=0.5\linewidth]{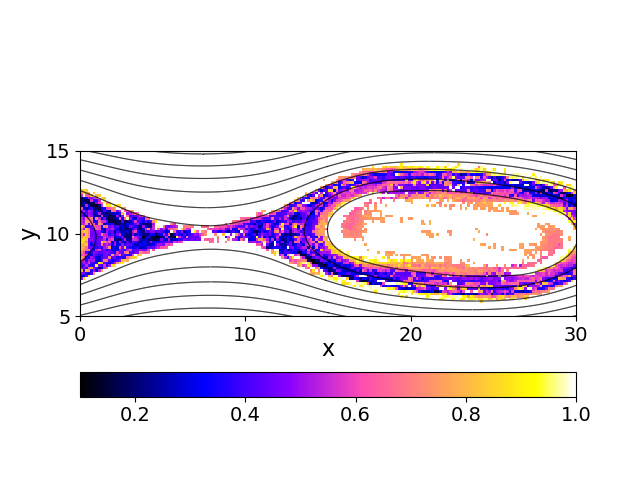}
  \includegraphics[trim={0.5cm 1.5cm 1.25cm 3.65cm},clip,width=0.5\linewidth]{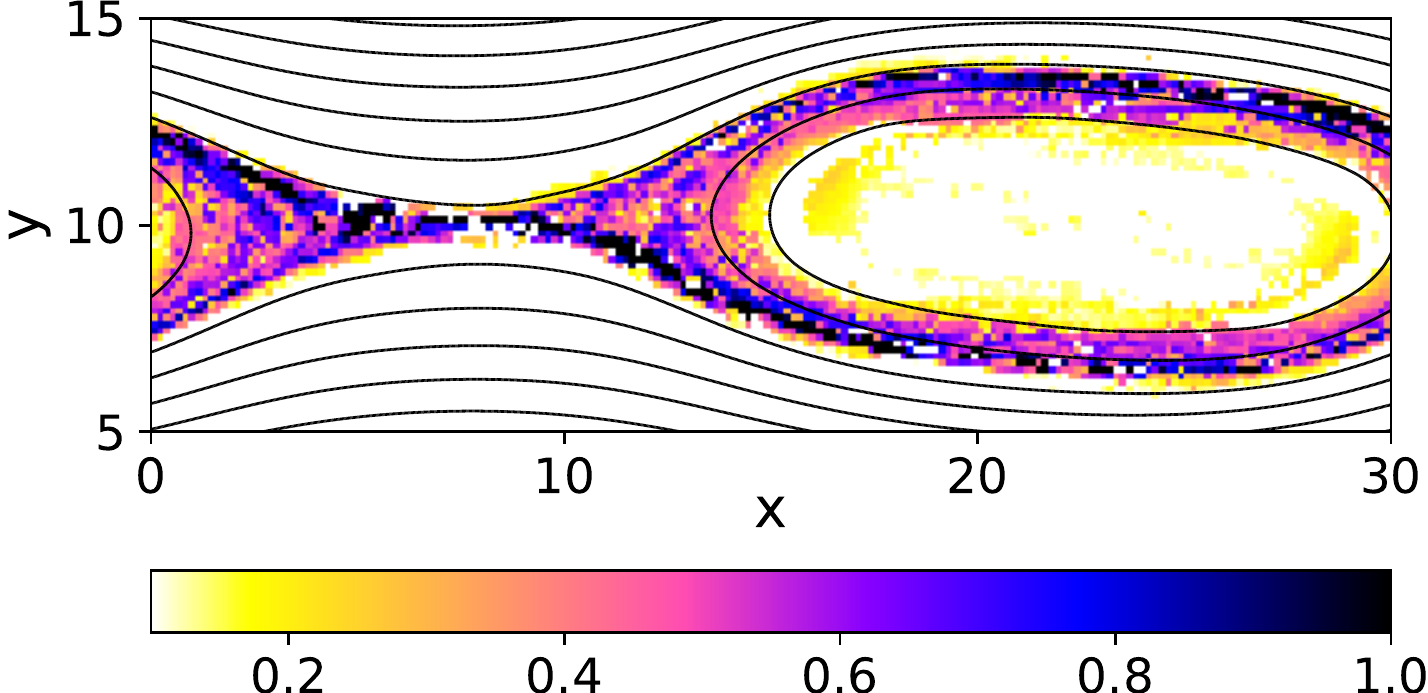}
}    
\caption{The left-hand column highlights the energy drop $E_{drop}$ defined by~\autoref{eq:energy_drop} and the right-hand column depicts the energy deviation $E_{dev}$ given by~\autoref{eq:energy_deviation}. Both quantities are presented at four different time steps, from top to bottom: $t=8,000$, $t=12,000$, $t=16,000$, and $t=20,000$.}
\label{fig:app_dh_result_strong_thermal} 
\end{figure}

\autoref{fig:app_dh_result_strong_thermal} presents the results for $E_{drop}$ and $E_{dev}$ and is less informative for the strong guide field. For instance, low $E_{drop}$ values around $0.4$ are observed for almost all the distributions with two or more components at $t=20,000$, meaning the presence of beams is more likely. As regards the energy deviation $E_{dev}$, high values above $0.8$ are observed from $t=12,000$ near the topological boundaries around the EDR, the outflow, and the separatrix. Thus, complex distributions are expected in these regions, which have also been highlighted by the measure of gyrotropy in~\autoref{fig:app_dh_result_strong}.

\section{Sensitivity to the resolution}
\label{app:resolution_sensitivity} 
Determining the number of components for the detection algorithm represents a not trivial model selection problem. In this paper, minimizing BIC has been chosen as the reference method. As stated in ~\autoref{eq:aic_bic}, the sample size directly influences BIC as the latter depends on $ln(n)$. Therefore, the detection algorithm is sensible to the number of particles provided to the GMM, encoded by the resolution chosen for the window defined in the section~\ref{sub:general_procedure}. The latter must be selected carefully to find a good trade off: a very broad window may mix several different particle populations, missing important physical scales while a very small window cannot reach a sufficient statistical convergence due to a very low number of particles.
\begin{figure*}[ht]
\gridline{
  \includegraphics[trim={0.5cm 3.7cm 1.3cm 3.85cm},clip,width=0.5\linewidth]{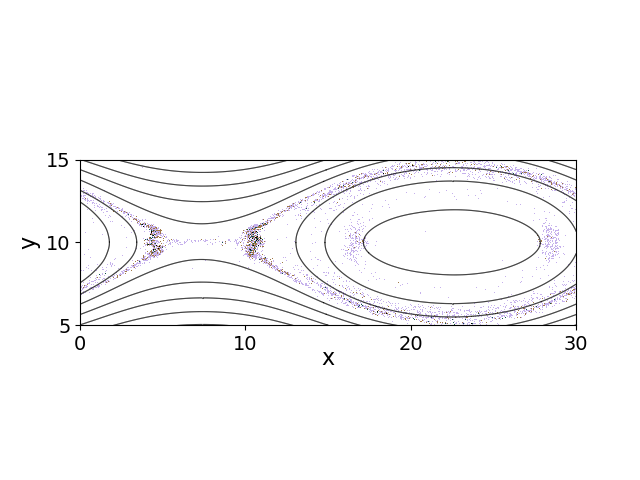}
}

\vspace{-0.35cm}

\gridline{
  \includegraphics[trim={0.5cm 3.7cm 1.3cm 3.85cm},clip,width=0.5\linewidth]{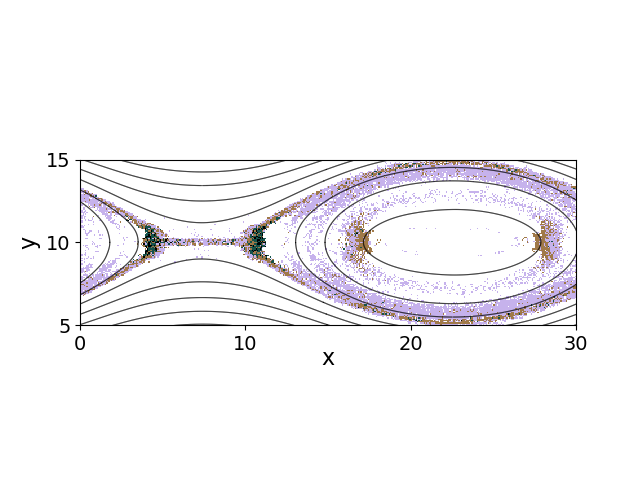}
}

\vspace{-0.35cm}

\gridline{
  \includegraphics[trim={0.5cm 3.7cm 1.3cm 3.85cm},clip,width=0.5\linewidth]{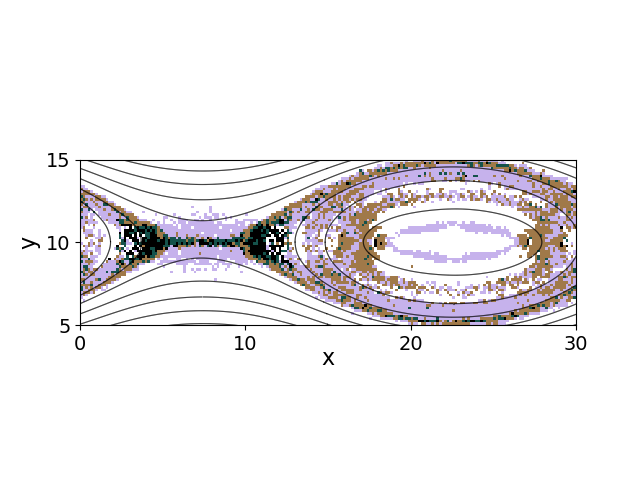}
}

\vspace{-0.35cm}

\gridline{
  \includegraphics[trim={0.5cm 1.5cm 1.3cm 3.65cm},clip,width=0.5\linewidth]{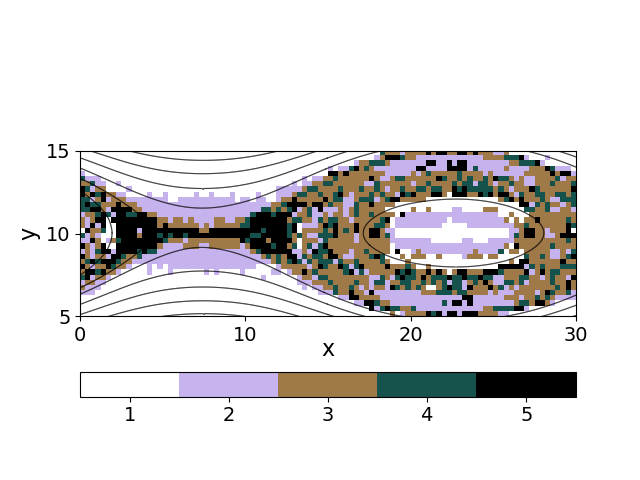}
}
\caption{Impact on the number of particles on the BIC optimization. Four different window length are depicted at $t=20,000$ from top to bottom: 1, 2, 4, and 8 cells.}
\label{fig:app_dh_result_weak_sensi} 
\end{figure*}

 For the double Harris sheet case with a weak guide field, a window of $4$ cells by $4$ cells has been selected, ensuring more than $1,000$ particles over the whole domain.~\autoref{fig:app_dh_result_weak_sensi} depicts the impact of the window length on the number of components provided by BIC minimization. Four lengths are investigated: $1$, $2$, $4$, and $8$. The smaller window ($r=1$ cell) barely detects the EDR and the topological boundaries. About a hundred particles are used to train each model, thus the mixture can miss important shapes and the underlying distributions are not necessary recovered. Typical structures start to be identified for $r=2$, where the EDR, outflow, O point, and separatrix regions show a significant size with a spatial correlation in term of number of components. Only the background region around the EDR is filtered out. Finally, the shapes identified for $r=4$ and $r=8$ look very similar, only few new distributions are observed, such as an intermediate region with $3$ components between the ion diffusion region and the EDR. Therefore,~\autoref{fig:app_dh_result_weak_sensi} illustrates perfectly the sensitivity of the algorithm and the BIC minimization to the window length and the number of particles. Even if characteristic structures of the reconnection are identified for each window length, from $1$ to $8$ cells, a minimum number of particles (here about $1,000$) is needed to ensure a proper statistical convergence while making sure different populations are not merged.

\section{Fixed number of components}
\label{app:fixed_number}
\begin{figure*}
\gridline{
  \includegraphics[trim={0.55cm 3.7cm 1.25cm 3.85cm},clip,width=0.5\linewidth]{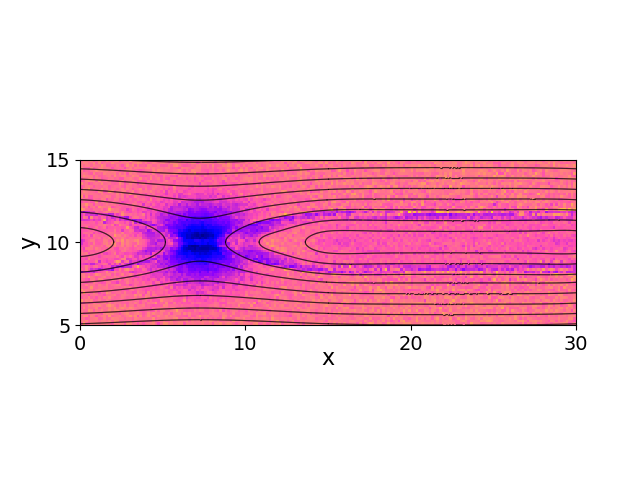}
  \includegraphics[trim={0.55cm 3.7cm 1.25cm 3.85cm},clip,width=0.5\linewidth]{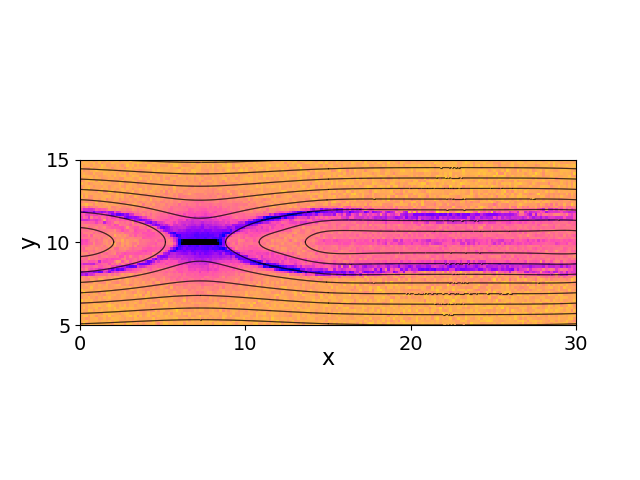}
}

\vspace{-0.35cm}

\gridline{
  \includegraphics[trim={0.55cm 3.7cm 1.25cm 3.85cm},clip,width=0.5\linewidth]{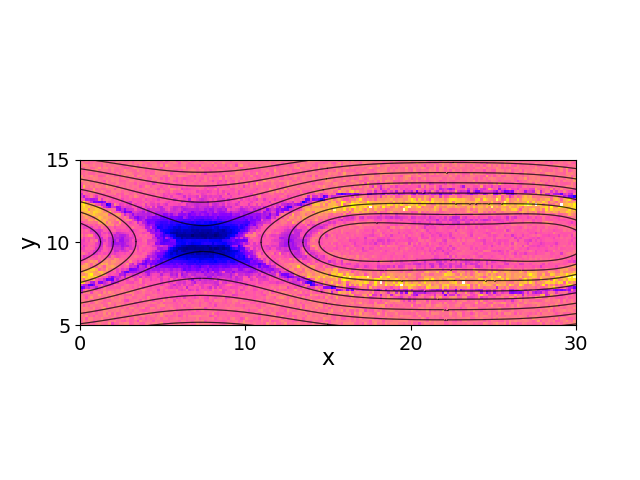}
  \includegraphics[trim={0.55cm 3.7cm 1.25cm 3.85cm},clip,width=0.5\linewidth]{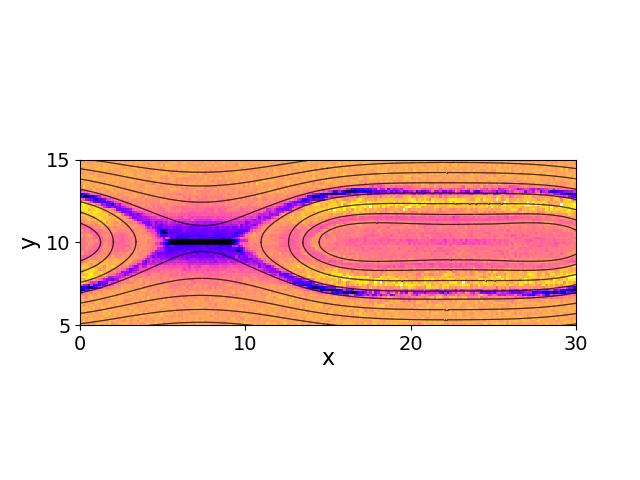}
}

\vspace{-0.35cm}

\gridline{
  \includegraphics[trim={0.55cm 3.7cm 1.25cm 3.85cm},clip,width=0.5\linewidth]{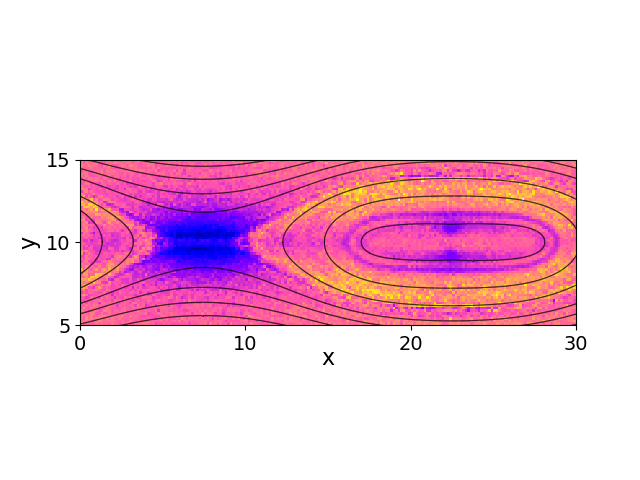}
  \includegraphics[trim={0.55cm 3.7cm 1.25cm 3.85cm},clip,width=0.5\linewidth]{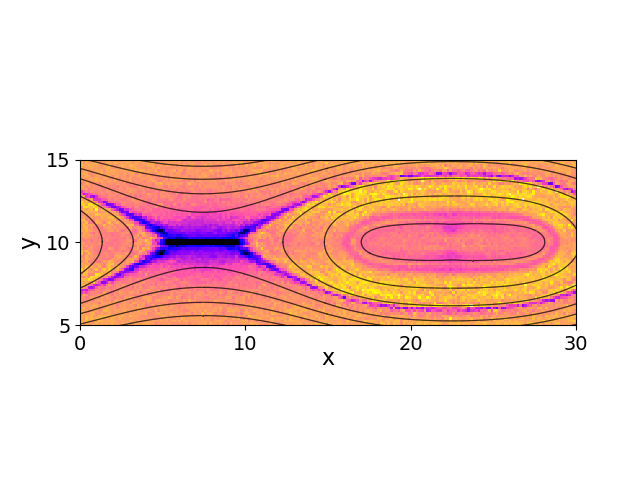}
}

\vspace{-0.35cm}

\gridline{
  \includegraphics[trim={0.55cm 1.5cm 1.25cm 3.65cm},clip,width=0.5\linewidth]{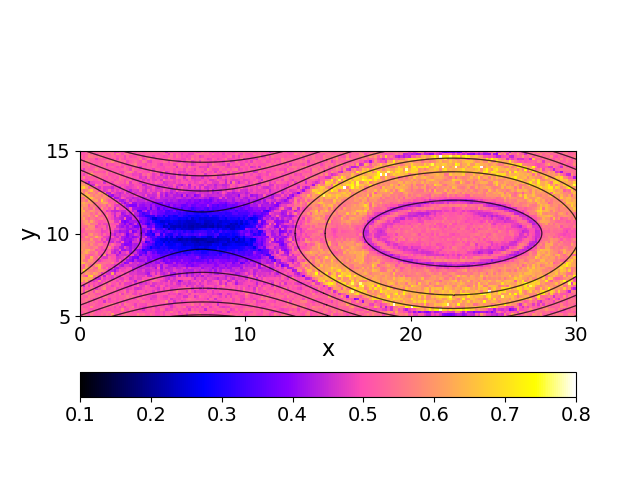}
  \includegraphics[trim={0.55cm 1.5cm 1.25cm 3.65cm},clip,width=0.5\linewidth]{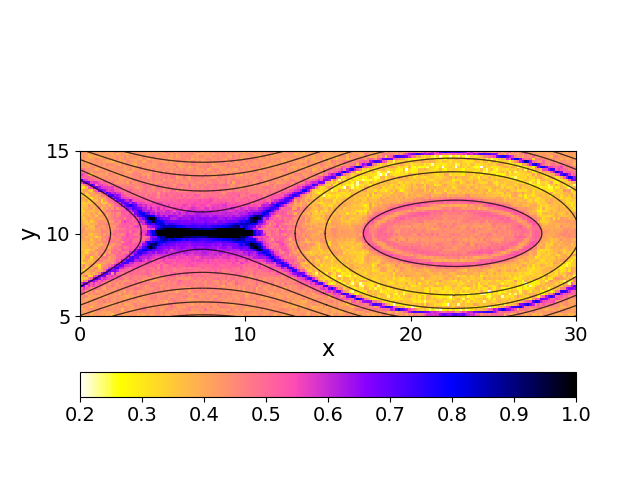}
}
\caption{Detection algorithm with a fixed number of $8$ components for the weak guide field case. The left-hand column highlights the energy drop $E_{drop}$ defined by~\autoref{eq:energy_drop} and the right-hand column depicts the energy deviation $E_{dev}$ given by~\autoref{eq:energy_deviation}. Both quantities are presented at four time steps, from top to bottom: $t=8,000$, $t=12,000$, $t=16,000$, and $t=20,000$.}
\label{fig:app_dh_result_weak_fixed} 
\end{figure*}

Here, a fixed and large number of components is imposed for the mixture models. It ensures a greater flexibility to the GMM as very complex distributions can be more easily described, reproducing nonparametric density estimation methods.~\autoref{fig:app_dh_result_weak_fixed} shows the energy drop $E_{drop}$ and the energy deviation $E_{dev}$. The structures identified by the algorithm are very close to the ones found with the BIC minimization depicted in~\autoref{fig:dh_result_weak_thermal}. For instance, the large background region surrounding the EDR is identified in both cases by the energy drop as well as the peak value of the EDR highlighted by $E_{drop}$. The only significant difference is observed for regions tagged with a single component by the BIC minimization. Distributions far upstream from the EDR have a noise level: around $0.5$ for $E_{drop}$ and around $0.4$ for $E_{dev}$ with a fixed number of components while BIC minimization provides values close to $1.0$ for $E_{drop}$ and almost zero values for $E_{dev}$. Similarly, the distributions within and around the O point show a significant noise level for a fixed number of components while this noise is filtered out by the automatic determination of the number of components. Therefore, BIC appears to be a good criterion to identify relevant statistical patterns in the data and to filtering out the noise in the distributions.

\clearpage
\bibliography{references} 



\end{document}